\documentclass[letterpaper, 12pt]{article}
\usepackage[textwidth=15.2cm,textheight=20cm]{geometry}
\usepackage{titling}
\usepackage{lipsum}
\pdfoutput=1
\usepackage[OT1]{fontenc}
\usepackage{float}
\usepackage{rotfloat}
\usepackage{bm}
\usepackage{amsthm, amsmath}
\usepackage{amstext}
\usepackage{amssymb}
\usepackage{graphicx}
\usepackage{esint}
\usepackage[authoryear]{natbib}
\usepackage{algorithm}
\newtheorem{cor}{Corollary}[section]
\newtheorem{lem}{Lemma}[subsection]
\usepackage{authblk}
\vspace{-80mm}
\title{A Bayesian changepoint methodology for high
	dimensional multivariate time series and space-time data: A study
of structural change using remotely sensed data.}

\newif\ifblinded

\blindedfalse

\ifblinded
  \author{}  
\else
   \author{Chris Strickland and Robert Burdett and
   Robert Denham and Robert Kohn and Kerrie Mengersen\footnote{c.strickland@unsw.edu.au, University of New South Wales,
 robert.burdett@qut.edu.au, Queensland University of Technology,
  robert.denham@qld.gov.au, Queensland Department of Science, Information Technology, Innovation and the Arts,
  r.kohn@unsw.edu.au, University of New South Wales, kerrie.mengersen@qut.edu.au, Queensland University of Technology}}
\fi

\begin{document}
\date{}
\maketitle




\begin{abstract}

A Bayesian approach is developed to analyze change points
in multivariate time series and space-time data. The methodology is
used to assess the impact of extended inundation on the ecosystem of
the Gulf Plains bioregion in northern Australia. The proposed approach
can be implemented for dynamic mixture models that have a conditionally
Gaussian state space representation. Details are given on how to efficiently
implement the algorithm for a general class of multivariate time series
and space-time models. This efficient implementation makes it feasible
to analyze high dimensional, but of realistic size, space-time data
sets because our approach can be appreciably faster, possibly millions
of times, than a standard implementation in such cases.
\end{abstract}

\section{Introduction}

The Gulf Plains bioregion of northern Australia is a large area of
tropical savanna on extensive alluvial plains and coastal areas
\citep{thackway1997bioregional}. The region experiences a monsoonal
climate with a winter dry season and a summer wet season, with the wet
season typically extending from October to April. A monsoon season
from December to March brings significant rainfall, often causing
flooding throughout the region. Flooding can be extensive, and
floodwaters can remain on pasture for several weeks.  In general, the
floodplains are resilient ecosystems, adapted to the wet and dry seasons,
with grasslands responding rapidly after the wet season. However,
periods of extended inundation can have an adverse lasting effect on
pasture, resulting in the death of grasses and seed bank.

The summer of 2008/2009 in this region experienced one of the most
severe floods on record, with widespread prolonged flooding from
January to March \citep{meteorology09}.  There was widespread reports
of death of grasses, and a lack of recovery the following year.  The
region is remote with limited infrastructure, so a remotely sensed
approach to monitor the extent and the timing of the event is
desirable.  In this paper, we use the Normalized Difference Vegetation
Index (NDVI) from NASA's Moderate Resolution Imaging Spectroradiometer
(MODIS) to determine the timing, effect size and recovery time
following an extended flood event. MODIS data is used extensively to
identify disturbance from time series data, though this typically uses
image difference techniques using only a small number of image dates
\citep[e.g.]{Jin2005462,Nielsen2007}, or uses univariate time series
models \citep[e.g.]{VerbesseltHyndmanNewnhamCulvenor,
verbesselt10:_phenol}.  The former approach is not suitable as
it ignores the time series structure present in the problem, while the
latter method ignores the information available through common trends,
as well as any spatial correlation present in the data. Unlike these
methods the approach presented in this paper, uses all of the data and
takes account of both the temporal and spatial correlation structure
in the data.

Specifically, we propose a method for analysing data with an unknown
number of changepoints that is applicable to high-dimensional
multivariate time series and space-time data. The detection of change
in time series and space-time data sets is of paramount importance in
many areas of statistics, and as a consequence it has been the focus
of much recent research; see, for example,
\citet{MajumdarGelfandBanerjee2004}, \citet{KoopPotter2007},
\citet{GiordaniKohn2008} and \citet{KoopPotter2009}.

Mixture innovation models, cast in a conditionally Gaussian state
space framework, provide an intuitive and flexible approach for
modeling non-linear effects, and furthermore they can be used for
models that allow for an unknown number of changepoints. The idea
behind this approach is to account for change by modeling the state
innovation using a mixture distribution. While approximate methods
have been developed for this class of models, see for example
\citet{HarrisonStephens1976} and \citet{SmithWest1983}, it has been
the advent of modern simulation methods that has facilitated the
development of exact methods. Early methods utilizing Markov chain
Monte Carlo (MCMC) include \citet{McCullochTsay1993} and
\citet{CarterKohn1994}. A drawback of these approaches is that
sampling the auxiliary discrete variables, used to define the mixture
on the innovations, is done conditionally on the states. This often
results in a poorly mixing MCMC sampler because of the high dependence
between the states and the auxiliary discrete variables. For the
univariate conditionally Gaussian state space model,
\citet{GerlachCarterKohn2000} propose an algorithm that generates the
auxiliary discrete variables in $O(n)$ operations, without
conditioning on the states, where $n$ is the sample size. This is an
important development because it overcomes the potentially high
dependence between the auxiliary discrete variables and the states
inherent in the algorithm of \citet{CarterKohn1994}.

Our article makes four major methodological contributions to the
Bayesian literature.  First, the methodology of
\citet{GerlachCarterKohn2000} is extended to multivariate
conditionally Gaussian state space models. Second, let $p$ be the
dimension of the observation vector in any time period.  Then for a
fairly general class of $p-$dimensional state space models, which apply
to both multivariate time series and space-time analysis,
we show how to sample from the posterior distribution of interest
using $O\left(pn\right)$ operations, rather than
$O\left(p^{3}n\right)$ operations, in the case of a naive
implementation of the algorithm.  We show that this results in a
increase in the speed that is of practical significance (possibly
millions of times faster) for the size of data that is of interest to
us. Third, the structure of the prior for the model and the MCMC
methodology allows us to average over the model space generated by the
common components in, possibly high dimensional, multivariate time
series and space-time models. This feature of our approach is very
important at a practical level, as well as being theoretically
attractive, particularly as the number of common components increase
in the specified model. Fourth, we propose a general approach for
sampling candidates in the latent state process of our model that is
efficient with respect to both computation and simulation. In
particular, we show how to draw in order $O(pn)$ operations any
parameter that enters the model only through the state transition
equation. Here, generating the parameter is done from its conditional
distribution with the states integrated out. We believe that each of
these contributions is necessary in producing a method, which contains
an adequately rich model structure and can be feasibly used to analyze
data sets of the size that are of interest to fields such as remote
sensing.

The new methodology is used to detect change in the NDVI
that is measured from the Moderate
Resolution Imaging Spectroradiometer satellite. The space-time data
set consists of nearly 18000 spatial locations at 268 time points.

\newgeometry{textwidth=15.2cm,textheight=20.5cm}
The article is organised as follows.
Section~\ref{sec:Condionally-Gaussian-Multivariat} describes the
conditionally Gaussian multivariate state space model and the new
sampling algorithm. Section~\ref{sec:Dynamic-Factor-Model} describes
the hierarchical multivariate time series and space-time model, its
efficient implementation, a computational comparison of the efficient
implementation and a naive implementation and a description of an MCMC
algorithm for the model. Section~\ref{sec:Analysis-of-Simulated}
demonstrates the methodology on simulated data.
Section~\ref{sec:Modelling-NDVI} utilizes the new methodology to
analyze structural change in the NDVI for the Gulf plains bioregion in
northern Australia.  data. Section~\ref{sec:Conclusions} summarizes
the article. All proofs of the results in the article are in the
appendix.

\section{Conditionally Gaussian Multivariate State Space
  Model\label{sec:Condionally-Gaussian-Multivariat}}

The observation vector $\bm{y}_{t}\in\mathbb{R}^{p},$ for
$t=1,2,\dots,n,$ for the conditionally Gaussian state space model
(CGSSM) is generated by
\begin{equation}
\bm{y}_{t}=\bm{g}_{t}+\bm{H}_{t}\bm{x}_{t}+\bm{G}_{t}\bm{e}_{t},\label{eq:meas}
\end{equation}
where $\bm{g}_{t}\in\mathbb{R}^{p}$, $\bm{H}_{t}\in\mathbb{R}^{p\times m}$
and $\bm{G}_{t}\in\mathbb{R}^{p\times p}$ are system matrices and
$\bm{e}_{t}\in\mathbb{R}^{p}$ is independently and normally distributed,
with mean $\bm{0}\in\mathbb{R}^{p}$ and a covariance $\bm{I}_{p},$
where $\bm{I}_{p}$ denotes an identity matrix of order $p$. The
state vector, $\bm{x}_{t}\in\mathbb{R}^{m},$ for $t=1,2,\dots,n-1,$
is generated by the difference equation
\begin{equation}
\bm{x}_{t+1}=\bm{h}_{t}+\bm{F}_{t}\bm{x}_{t}+\bm{\Gamma}_{t}\bm{u}_{t},\label{eq:state}
\end{equation}
with $\bm{h}_{t}\in\mathbb{R}^{m},$ the transition matrix
$\bm{F}_{t}\in\mathbb{R}^{m\times m},$
$\bm{\Gamma}_{t}\in\mathbb{R}^{m\times r}$ and the disturbance vector
$\bm{u}_{t}\in\mathbb{R}^{r}$ is defined to be serially uncorrelated
and normally distributed with a mean $\bm{0}\in\mathbb{R}^{r}$ and a
covariance matrix $\bm{I}_{r}$. The system matrices in (\ref{eq:meas})
and (\ref{eq:state}) are functions of the unknown parameters
$\bm{\omega}\in\mathbb{R}^{l}$ and also depend on a sequence of
discrete random variables $\bm{K}_{t}\in\mathbb{R}^{s},$ which can be
used to model non-linear effects in an intuitive manner. The state space
model is completed by specifying the distribution of the initial state
$\bm{x}_{1}$ as
\begin{equation}
\bm{x}_{1}\sim N\left(\bm{m}_{1},\bm{V}_{1}\right),\label{eq:init_state}
\end{equation}
with mean $\bm{m}_{1}\in\mathbb{R}^{m}$ and covariance
$\bm{V}_{1}\in\mathbb{R}^{m\times m}$. For notational convenience
throughout, denote
$\bm{x}=\left(\bm{x}_{1},\bm{x}_{2},\dots,\bm{x}_{n}\right)$ and
$\bm{x}^{s:t}=\left(\bm{x}_{s},\bm{x}_{s+1},\dots,\bm{x}_{t}\right),$
where this convention extends to any vector or matrix.

\subsection{Estimation}

Algorithm~\ref{alg:MCMC} provides a general way of estimating the
model described in (\ref{eq:meas})--(\ref{eq:init_state}).

\begin{algorithm}[H]
\begin{enumerate}
\item Sample $\bm{K}^{\left(j\right)}$ from $p\left(\bm{K}|\bm{y},\bm{\omega}^{\left(j-1\right)}\right)$
\item Sample $\bm{x}^{\left(j\right)}$ from
  $p\left(\bm{x}|\bm{y},\bm{K}^{\left(j\right)},\bm{\omega}^{\left(j-1\right)}\right)$.
\item Sample $\bm{\omega}^{\left(j\right)}$ from $p\left(\bm{\omega}|\bm{y},\bm{x}^{\left(j\right)},\bm{K}^{\left(j\right)}\right)$.
\end{enumerate}
\caption{\label{alg:MCMC}}
\end{algorithm}

Step 3 of Algorithm~\ref{alg:MCMC} is model specific. Steps 1
and 2 can be completed using algorithms that are applicable to the
general state space model. Specifically, in Step 1, $\bm{K}$ is sampled
from $p\left(\bm{K}|\bm{y},\bm{\omega}\right)$ by sampling each
$\bm{K}_{t},$ for $t=1,2,\dots,n,$ from
$p\left(\bm{K}_{t}|\bm{y},\bm{K}_{s\ne t}\right).$ The algorithm used
in this computation is described below. In Step 2, $\bm{x}$ is drawn
from $p\left(\bm{x}|\bm{y},\bm{K},\bm{\omega}\right).$ This step,
which involves sampling the state from its full conditional posterior
distribution, can be achieved using any of the algorithms
developed by \citet{CarterKohn1994}, \citet{FruhwirthSchnatter1994},
\citet{deJongShepard1995}, \citet{DurbinKoopman2002} or
\citet{StricklandTurnerDenhamMengerson2009}.  Sampling $\bm{K}$ is
the most difficult step, and is achieved through a generalization of
results presented by \citet{GerlachCarterKohn2000} who propose an algorithm to efficiently
sample $\bm{K},$ for the univariate state space model, i.e.\ when
$\bm{y}_{t}$ is a scalar. Their contribution is to show that $\bm{K}$ can
be sampled in $O\left(n\right)$ operations, without needing to
condition on the states $\bm{x}.$ The idea builds on the relation
\begin{eqnarray}
p\left(\bm{K}_{t}|\bm{y},\bm{K}_{s\ne t},\bm{\omega}\right) & \propto & p\left(\bm{y}|\bm{K},\bm{\omega}\right)\times p\left(\bm{K}_{t}|\bm{K}_{s\ne t},
	\bm{\omega}\right)\nonumber \\
 & \propto & p\left(\bm{y}^{t+1:n}|\bm{y}^{1:t},\bm{K},\bm{\omega}\right)\times p\left(\bm{y}_{t}|\bm{y}^{1,t-1},\bm{K}^{1:t},\bm{\omega}\right)\times\label{eq:posterior_K}\\
 &  & p\left(\bm{K}_{t}|\bm{K}_{s\ne t},\bm{\omega}\right),
\end{eqnarray}
where the term $p\left(\bm{K}_{t}|\bm{K}_{s\ne t},\bm{\omega}\right)$
is obtained from the prior, and may depend on unknown parameters,
the term $p\left(\bm{y}_{t}|\bm{y}^{1,t-1},\bm{K}^{1:t},\bm{\omega}\right)$
is obtained using one step of the Kalman filter and the term $p\left(\bm{y}^{t+1:n}|\bm{y}^{1:t},\bm{K},\bm{\omega}\right)$
is obtained by one forward step after initially doing a set of backward
recursions.

Given Lemmas \ref{Lemma1}, \ref{Lemma2} and \ref{Lemma3} (see the appendix),  an algorithm for sampling $\bm{K}$ from $p\left(\bm{K}|\bm{y},\bm{\theta}\right)$
is defined as follows:

\begin{algorithm}[H]
\begin{enumerate}
\item Given the current value of $\bm{K},$ for $t=n-1,n-2,\dots,1,$ compute
$\bm{\mu}_{t}$ and $\bm{\Omega}_{t}$ using the recursion given in
\ref{Lemma1}.
\item For $t=1,2,\dots,n,$

\begin{enumerate}
\item compute $p\left(\bm{y}_{t}|\bm{y}^{1:t-1},\bm{K}^{1:t},\bm{\omega}\right)$
using \ref{Lemma2},
\item compute $p\left(\bm{y}^{t+1:n}|\bm{y}^{1:t},\bm{K},\bm{\omega}\right)$
using \ref{Lemma3},
\item for all values of $\bm{K}_{t}$ compute $p\left(\bm{K}_{t}|\bm{y},\bm{\omega}\right)$
and form a probability mass function and use it to sample $\bm{K}_{t}.$
\item re-run one step of the Kalman filter, defined in \ref{Lemma2}, based
on the sample $\bm{K}_{t}.$ \caption{Algorithm to sample $\bm{K}.$ \label{alg:Sample_K}}
\end{enumerate}
\end{enumerate}
\end{algorithm}

\section{Hierarchical Time Series and Space-Time Modeling\label{sec:Dynamic-Factor-Model}}

We consider a general modeling framework that applies to both
multivariate time series and space-time analysis. In particular, the
observation equation at time $t,$ for $t=1,2,\dots,n,$ for the
observations, $\bm{y}_{t}\in\mathbb{R}^{p},$ is
\begin{equation}
\bm{y}_{t}=\bm{\Theta}\bm{f}_{t}+\bm{\bm{e}}_{t},\label{eq:DFM}
\end{equation}
where $\bm{\Theta}\in\mathbb{R}^{p\times k}$ is a matrix of basis
functions, which is possibly spatially referenced,
$\bm{f}_{t}\in\mathbb{R}^{k}$ is a vector of common components and
$\bm{e}_{t}\in\mathbb{R}^{p}$ is a vector of serially uncorrelated,
normally distributed disturbances, with diagonal covariance matrices
$\bm{\Sigma}_{t}\in\mathbb{R}^{p\times p}$.  The matrix of basis
functions, $\bm{\Theta},$ in the case of multivariate time series
analysis is typically taken as unknown and the model in (\ref{eq:DFM})
is commonly referred to as a dynamic factor model (DFM). In space-time
analysis a wide variety of basis functions have been used in its
specification, including empirical orthogonal functions EOF,
Fourier and wavelet bases, amongst many other methods; see
\citet{CressieWikle2011} for a thorough review. When $\bm{\Theta}$ is
unknown, it is assumed to be a function of a vector of parameters,
$\bm{\kappa},$ which needs to be estimated. We specify the structure
of $\bm{\Theta}$ when applying the model in
Sections~\ref{sec:Analysis-of-Simulated}
and~\ref{sec:Modelling-NDVI}. Regardless of the specification of the
matrix of basis functions, its purpose is the same: to provide a
mapping between the high dimensional set of observations and a low
dimensional system that aims to capture the dynamic characteristics in
the data generating process. This method of dimension reduction is
necessary computationally and practically sensible. For example, in
remotely sensed data one may expect that observations in woodlands
might exhibit a common temporal signature, and data points in
grasslands exhibit a different but also common temporal signature. Our
models can take advantage of such features present in the
data. Typically the common terms, $\bm{f}_{t},$ are sums of a number
of components such as trend, regression and autoregressive
components. To estimate the model in a state space framework, it is
convenient to define $\bm{f}_{t}=\bm{\Phi}\bm{x}_{t},$ where
$\bm{\Phi}\in\mathbb{R}^{k\times m}$ is a selection matrix that is
used so that the model in (\ref{eq:DFM}) can be written in state space
form. It follows that,
\begin{eqnarray}
\bm{y}_{t} & = & \bm{\Theta}\bm{\Phi}\bm{x}_{t}+\bm{e}_{t},\label{eq:meas_DFM}\\
\bm{x}_{t+1} & = & \bm{W}_{t}\bm{\beta}+\bm{F}_{t}\bm{x}_{t}+\bm{\Lambda}_{t}\bm{v}_{t},\label{eq:state_DFM}
\end{eqnarray}
where $\bm{W}_{t}\in\mathbb{R}^{m\times k^{R}}$ is a matrix of regressors,
$\bm{\beta}\in\mathbb{R}^{k^{R}}$ is a vector of regression coefficients,
$\bm{\Lambda}_{t}\in\mathbb{R}^{m\times r}$ and $\bm{v}_{t}\in\mathbb{R}^{r}$
is a random vector that is normally distributed with a covariance
matrix $\bm{I}_{m}$. It is immediately apparent that (\ref{eq:meas_DFM})
and $\left(\ref{eq:state_DFM}\right)$ can be expressed as the CGSSM
in (\ref{eq:meas}) and (\ref{eq:state}), by defining $\bm{H}_{t}=\bm{\Theta}\bm{\Phi},$
$\bm{G}_{t}=\bm{\Sigma}_{t}^{\frac{1}{2}},$ $\bm{h}_{t}=\bm{W}_{t}\bm{\beta}$
and $\bm{\Gamma}_{t}=\bm{\Lambda}_{t}.$ Note that for certain classes
of basis functions, it may be necessary to impose identification restrictions
on the model.

\subsection{The Multiple Change Point Problem}

Modeling multiple change points, using a variation of
Algorithm~\ref{alg:MCMC}, is accomplished by defining
$\bm{\Lambda}_{t}$ to be a function of $\bm{K}_{t}.$ Specifying a
model to handle changepoints in this way is simple and flexible. For
example, we can specify each common component to consist of an
autoregressive process and a level that allows for shifts in the
conditional mean and slope of the process. This is achieved, for
$i\in\mathbb{N}^{k},$ where $\mathbb{N}^{k}$ denotes the set $\left\{
  1,2,\dots,k\right\} ,$ by defining
\begin{equation}
f_{i,t}=\psi_{i,t}+\mu_{i,t}+\bm{w}_{t-1}^{T}\bm{\beta}_{i},\label{eq:common_factor_i}
\end{equation}
where $\psi_{i,t}$ is an autoregressive cyclical process, $\mu_{i,t}$
is the level, $\bm{w}_{t}\in\mathbb{R}^{k^{r}}$ is a vector of
regressors and $\bm{\beta}_{i}\in\mathbb{R}^{k^{r}}$ is a vector of
regression coefficients. Specifically, we define $\psi_{i,t}$ as a
damped stochastic cycle, such that
\[
\psi_{i,t+1}=\rho_{i}\left(\cos\left(\lambda_{i}\right)\psi_{i,t}+\sin\left(\lambda_{i}\right)\psi_{i,t}^{*}\right)+\sigma_{f,i}\zeta_{i,t},
\]
where $\rho_{i}$ is a persistence parameter, $\lambda_{i}$ is a
hyperparameter that defines the period of the cycle, $\psi_{i,t}^{*}$
is an auxiliary variable defined by
\[
\psi_{i,t+1}^{*}=\rho_{i}\left(\cos\left(\lambda_{i}\right)\psi_{i,t}^{*}-\sin\left(\lambda_{i}\right)\psi_{i,t}\right)+\sigma_{f,i}\zeta_{i,t}^{*},
\]
$\sigma_{f,i}$ is a scale parameter and $\zeta_{i,t}$ and
$\zeta_{i,t}^{*}$ are standard normal random variables. The stochastic
cycle reverts to a standard first order autoregressive process when
$\lambda_{i}=0;$ for further details on stochastic cycles, see
\citet{Harvey1989}.  The cycle is used to capture seasonal effects in
the analysis in this paper. The level $\mu_{i,t}$ is modeled as
\begin{align*}
\mu_{i,t+1} & =  \mu_{i,t}+\delta_{i,t}+\sigma_{f,i}K_{i,t}^{\mu}\xi_{i,t},&
\delta_{i,t+1} & =  \delta_{i,t}+\sigma_{f,i}K_{i,t}^{\delta}\chi_{i,t}
\end{align*}
where $\delta_{i,t}$ captures the slope for the $i^{th}$ common
component, $K_{i,t}^{\mu}$ is a discrete random variable that is used
to accommodate change in the level, $\xi_{i,t}$ and $\chi_{i,t}$ are
independent standard normal random variables; $K_{i,t}^{\delta}$ is a
discrete random variable that is used to model changes in the slope.

For $i\in\mathbb{N}^{k},$ the prior for $\rho_{i}$ is a beta
distribution, $\mathcal{B}\left(\alpha_{\rho},\beta_{\rho}\right)$,
which ensures that $\psi_{t,i}$ is a stationary process, with a
positive autocorrelation function. The prior for $\lambda_{i}$ is a
\emph{stretched }beta distribution,
$\mathcal{B}^{\left(a,b\right)}\left(\alpha_{\lambda},\beta_{\lambda}\right),$
where
$\mathcal{B}^{(a,b)}\left(\alpha_{\lambda},\beta_{\lambda}\right)$ is
a beta distribution that has been translated and stretched over the
open set $\left(a,b\right)$, i.e., if $\zeta_i\sim
\mathcal{B}\left(\alpha_{\lambda},\beta_{\lambda}\right),$ then
$\lambda_i=a+\zeta_i(b-a)$. Let $\beta_{i,j}$ be the $j^{th}$ element
of $\bm{\beta}_{i}.$ Then, the $\beta_{i,j}$ are \emph{a priori}
independent, i.e.,
$p\left(\bm{\beta}_{i}\right)=p\left(\beta_{i,1}\right)\times
p$$\left(\beta_{i,2}\right)\times\cdots\times
p\left(\beta_{i,k^{r}}\right)$ and \[
p\left(\beta_{i,j}\right)=\left(1-\varpi_{i,j}\right)\delta_{0}\left(\beta_{i,j}\right)+\varpi_{i,j}\mathcal{N}\left(0,\sigma_{\beta}^{2}\right),
\] where $\varpi_{i,j}\in\left\{ 0,1\right\} $ is a Bernoulli
auxiliary random variable, such that
$\varpi_{i.j}\sim\mathcal{B}ern\left(1,p_{\varpi}\right).$ For
$i\in\mathbb{N}^{k},$ the prior for $\sigma_{f,i}$ is an inverted
gamma distribution, i.e., $\sigma_{f,i}\sim
IG\left(\nu_{f\sigma}/2,s_{f\sigma}/2\right),$ where
$\nu_{f_{\sigma}}$ is the degrees of freedom parameter and
$s_{f\sigma}$ is the scale parameter.

In our article, the discrete random variables for the $i^{th}$ common
component, $K_{i,t}^{\mu}\in\left\{
  0,\eta_{1,i}^{\mu},\eta_{2,i}^{\mu}\right\} $ and
$K_{i,t}^{\delta}\in\left\{
  0,\eta_{1,i}^{\delta},\eta_{2,i}^{\delta}\right\} $ are used to
capture changes in the level and slope, respectively.  The elements of
$\bm{\eta}=\left(\eta_{1,i}^{\mu},\eta_{2,i}^{\mu},\eta_{1,i}^{\delta},\eta_{2,i}^{\delta}\right)$
are assumed independent \emph{a priori} with prior distributions that
are inverted gamma, where for $j\in\left\{ 1,2\right\} ,$
$\eta_{j,i}^{\mu}\sim IG\left(\nu_{j}^{\mu}/2,s_{j}^{\mu}/2\right)$
and $\eta_{j}^{\delta}\sim
IG\left(\nu_{j}^{\delta}/2,s_{j}^{\delta}/2\right).$ It is assumed
that $\bm{K}_{t}=\left\{ K_{i,t}^{\mu},K_{i,t}^{\delta}\right\}
_{i\in\mathbb{N}^{k}}$ is multinomial, where at each time point only
one change point (in either the level or the slope) is allowed for all
$i.$ While we can specify a more general prior, we found that this
prior works well for the data sets we have analyzed. The multinomial
prior distribution for $\bm{K}_{t}$ is defined assuming that we are in
the null state, i.e. $K=0|K\in\bm{K}_{t},$ with probability $1-\pi.$
It is further assumed that $\bm{K}_{t}$ takes any other possible
values with equal probability.

\begin{table}[h]
\centering
\begin{tabular}{cccccccccc}
\hline
$K_{1,t}^{\mu}$  & 0  & $\eta_{1,1}^{\mu}$  & $\eta_{2,1}^{\mu}$  & 0  & 0  & 0  & 0  & 0  & 0\tabularnewline
$K_{2,t}^{\mu}$  & 0  & 0  & 0  & $\eta_{1,2}^{\mu}$  & $\eta_{2,2}^{\mu}$  & 0  & 0  & 0  & 0\tabularnewline
$K_{1,t}^{\delta}$  & 0  & 0  & 0  & 0  & 0  & $\eta_{1,1}^{\delta}$  & $\eta_{2,1}^{\delta}$  & 0  & 0\tabularnewline
$K_{2,t}^{\delta}$  & 0  & 0  & 0  & 0  & 0  & 0  & 0  & $\eta_{1,2}^{\delta}$  & $\eta_{2,2}^{\delta}$\tabularnewline
$p\left(\bm{K}_{t}\right)$  & $\left(1-\pi\right)$  & $\frac{\pi}{8}$  & $\frac{\pi}{8}$  & $\frac{\pi}{8}$  & $\frac{\pi}{8}$  & $\frac{\pi}{8}$  & $\frac{\pi}{8}$  & $\frac{\pi}{8}$  & $\frac{\pi}{8}$\tabularnewline
\hline
\end{tabular}\caption{\label{tab:Prior_K}Prior distribution for $\bm{K}_{t}$.}
\end{table}

For example, Table~\ref{tab:Prior_K} illustrates the case for the two
component model. The first four rows of the table capture the possible
values for $\bm{K}_{t},$ while the bottom row reports the prior
probability of being in each state. For example, the second column of
the table shows that the probability of being in the null state is
$\left(1-\pi\right),$ while the third column states that
$K_{1,t}^{\mu}=\eta_{1,1}^{\mu},K_{2,t}^{\mu}=K_{1,t}^{\delta}=K_{2,t}^{\delta}=0,$
with probability $\frac{\pi}{8}.$

The model for the $i^{th}$ common component in
(\ref{eq:common_factor_i}) nests many models of interest. In
particular, as this approach can be used to account for an unknown
number of changepoints, it nests every possibility from the case of no
changepoints, i.e. $K_{i,t}^{\mu}=K_{i,t}^{\delta}=0$ for all $t$ so
that the model for $f_{i}$ is the cycle plus regression component, to
the case of a changepoint in either the mean or slope at every
observation. This approach, which determines the model as part of the
estimation avoids the need to specify a model for each common
component and is of increasing practical importance as the
number of common components grows. A theoretical attraction of this
approach is that it averages over the model space of common factors
and the $\bm{K}_{t}$ discrete variables to correctly account for
uncertainty in the model.

The hierarchical model of interest, in which the common components are
specified according to (\ref{eq:common_factor_i}), can be formulated
following (\ref{eq:meas_DFM}) and (\ref{eq:state_DFM}), by defining
$\bm{x}_{t}=\left[\begin{array}{cccc} \bm{x}_{1,t}^{T} &
    \bm{x}_{2,t}^{T} & \cdots &
    \bm{x}_{k,t}^{T}\end{array}\right]^{T},$ where
$\bm{x}_{i,t}=\left[\begin{array}{cccc} \tilde{\psi}_{i,t} &
    \psi_{i,t}^{*} & \mu_{i,t} & \delta_{i,t}\end{array}\right]^{T},$
with $\tilde{\psi}_{i,t}=\psi_{i,t}-\bm{w}_{t-1}\bm{\beta}_{i}$ and
$\bm{\Phi}=\text{diag}\left(\bm{\phi}_{1}^{T},\bm{\phi}_{2}^{T},\dots,\bm{\phi}_{k}^{T}\right),$
with
$\bm{\phi}_{1}=\bm{\phi}_{2}=\cdots=\bm{\phi}_{k}=\left[\begin{array}{cccc}
    1 & 0 & 1 & 0\end{array}\right].$ Furthermore, the state
transition matrix
$\bm{F}_{t}=\text{diag}\left(\bm{F}_{1,t},\bm{F}_{2,t},\dots,\bm{F}_{k,t}\right),$
with $\bm{F}_{i,t}$ a $4 \times 4 $ block diagonal matrix, with
$(\bm{F}_{i,t})_{11} = (\bm{F}_{i,t})_{22} = \rho_i \cos(\lambda) , (\bm{F}_{i,t})_{12} = - (\bm{F}_{i,t})_{21} = \rho_i \sin(\lambda) $
and $(\bm{F}_{i,t})_{33} = (\bm{F}_{i,t})_{34} = {\bm{F}_{i,t}}_{44} = 1$; the rest of the elements of $\bm{F}_{i,t}$ are zero.
The system matrix $\bm{\Lambda}_{t}=\text{diag}\left(\bm{\Lambda}_{1,t},\bm{\Lambda}_{2,t},\dots,\bm{\Lambda}_{k,t}\right),$
 with $\bm{\Lambda}_{i,t}=\text{diag}\left(1, 1, K_{1,t}^{\mu},
 K_{2,t}^{\delta}\right)$.
 The regressors are formulated such that for $t=1,2,\dots,n-1,$$\bm{W}_{t}=\text{diag}\left(\bm{W}_{t,1},\bm{W}_{t,2},\dots, \bm{W}_{t,k}\right),$
 where $\bm{W}_{t,i}$ is a $(4\times 2)$ matrix, with $(\bm{W}_{t,i})_{11}=\bm{w}_{t}-\rho_{i}\cos\left(\lambda_{i}\right)\bm{w}_{t-1}$,
 $(\bm{W}_{t,i})_{21}=\rho_{i}\sin\left(\lambda_{i}\right)\bm{w}_{t-1}$, and where the rest of the elements in $\bm{W}_{t,i}=0$;
 $\bm{W}_0$ is a $(4\times 2)$ matrix, with $(\bm{W}_0)_{11}=\omega_0$, $(\bm{W}_0)_{31}=1$; the rest of the elements in $\bm{W}_{0}=0$.
 Note that $\bm{\beta}=\left(\bm{\beta}_{1}^{T},\mu_{1,1},\bm{\beta}_{2}^{T},\mu_{2,1},\dots,\bm{\beta}_{k}^{T},\mu_{k,1}\right).$

\subsection{Efficient Estimation}

Sampling $\bm{K}=\left\{ \bm{K}_{t},\, t=1,2,\dots,n\right\} $ for the
hierarchical model in (\ref{eq:meas_DFM}) and (\ref{eq:state_DFM})
using Algorithm~\ref{alg:Sample_K} is straightforward but very
inefficient, as it involves $O\left(p^3n\right)$ operations. The following Lemma
(see Appendix \ref{ProofLemma4}
for a proof) shows how to sample
$\bm{K}_{t}$ from $p\left(\bm{K}_{t}|\bm{y},\bm{K}_{s\ne
    t},\bm{\omega}\right)$ far more efficiently.
\begin{lem}
\label{Lemma4}
  Suppose that in the hierarchical model in (\ref{eq:meas_DFM}) and
  (\ref{eq:state_DFM}), $\bm{K}$ enters only through the state
  equation. Then it is possible to sample $\bm{K}$ using
  $p\left(\bm{K}_{t}|\bm{y},\bm{K}_{s\ne t},\bm{\omega}\right)$ by
  applying Algorithm~\ref{alg:Sample_K} to the transformed state space
  model
\begin{equation}
\bm{y}_{t}^{L}=\bm{\Psi}\bm{x}_{t}+\bm{e}_{t}^{L},\label{eq:transformed_meas}
\end{equation}
where
$\bm{y}_{t}^{L}=\left(\bm{\Theta}^{T}\bm{\Sigma}_{t}^{-1}\bm{\Theta}\right)^{-1}\bm{\Theta}^{T}\bm{\Sigma}^{-1}\bm{y}_{t},$
$\bm{e}_{t}\sim\mathcal{N}\left(\bm{0},\bm{\Sigma}_{t}^{L}\right),$
with
$\bm{\Sigma}_{t}^{L}=\left(\bm{\Theta}^{T}\bm{\Sigma}_{t}^{-1}\bm{\Theta}\right)^{-1}$
and the state equation remains the same as (\ref{eq:state_DFM}).
\end{lem}
The transformation in (\ref{eq:transformed_meas}) is motivated by
\citet{JungbackerKoopman}, who suggest using the same transformation
in sampling the state, $\bm{x},$ from its full conditional posterior
distribution, for the case of the DFM\@. Jungbacker and Koopman also
show how to modify the likelihood to correct for the transformation.
The lemma shows that it is unnecessary to modify Lemma~\ref{Lemma4} to sample
$\bm{K}$. The computational savings that arise from applying
Algorithm~\ref{alg:Sample_K} are dramatic, if the number of time
series, $p,$ is large, which is common in space-time analysis; see for
example \citet{StricklandSimpsonTurnerDenhamMengerson2011}, where $p$
is close to one thousand, or perhaps even more dramatically in the
analysis in Section~\ref{sec:Modelling-NDVI} where $p$ is close to
18000.  Specifically, if Algorithm~\ref{alg:Sample_K} is implemented
on the model that has not be transformed then $O\left(p^{3}n\right)$
operations are required. However, for the $k$-dimensional state
space model in (\ref{eq:transformed_meas}) and (\ref{eq:state_DFM})
only $O\left(m^{3}n\right)$ operations are required, where the
dimension of the state, $m,$ is typically equal to or slightly larger
than $k$, and $k\ll p.$ As such, the main cost of using the
transformed model in (\ref{eq:transformed_meas}) typically comes from
the computation of the transform, which requires $O\left(pn\right)$
operations.
\begin{cor}
\label{Cor1}
Suppose that $\bm{\omega}_{s}$ is a vector of parameters that only
appears in the state equation. Then the density
\begin{equation}
p\left(\bm{\omega}_{s}|\bm{y},\bm{\Theta},\bm{\Sigma},\bm{K}\right)\propto p\left(\bm{y}_{L}|,\bm{K},\bm{\Sigma}_{L}\right)p\left(\bm{\omega}_{s}\right),\label{eq:Lemma_5}
\end{equation}
where $p\left(\bm{y}_{L}|,\bm{K},\bm{\Sigma}_{L}\right)=\int p\left(\bm{y}_{L},\bm{x}|\bm{\Sigma}_{L},\bm{K}\right)d\bm{x}$
can be computed by applying the Kalman filter to the lower dimensional
SSM in (\ref{eq:Transformed_DFM}) and (\ref{eq:state_DFM}).
\end{cor}
Corollary~\ref{Cor1} is particularly important in constructing efficient
sampling schemes for large data sets. Its proof follows from that of
Lemma~\ref{Lemma4}. When using MCMC to analyze large data sets it is important
that each component of the MCMC algorithm can be calculated in a
computationally efficient manner, and furthermore induce as little
correlation as possible in the Markov chain. In some sense the chain
is only as strong as its weakest link and thus even if just one
component of the MCMC algorithm is inefficient then this can be
enough, at least for the large data case, to render the MCMC algorithm
impractical. It is straightforward to implement an adaptive RWMH
algorithm, based on (\ref{eq:Lemma_5}), to sample any of the
hyperparameters in the state, where the form of (\ref{eq:Lemma_5})
ensures that we are sampling the parameter of interest marginal of
$\bm{x}.$ Sampling state hyperparameters, marginal of the state, is
shown to significantly improve the simulation efficiency of the
resultant estimates in \citet{KimShephardChib1998} and
\citet{StricklandTurnerDenhamMengerson2009}. While we can
alternatively compute
$p\left(\bm{\omega}_{s}|\bm{y},\bm{\Theta},\bm{\Sigma},\bm{K}\right)\propto
p\left(\bm{y}|,\bm{K},\bm{\Sigma}\right)p\left(\bm{\omega}_{s}\right),$
by applying the Kalman filter to the full model in (\ref{eq:meas_DFM})
and (\ref{eq:state_DFM}), this would be practically infeasible for
large data sets.

\subsubsection{Computational Comparison}

To illustrate the practical importance of Lemma~\ref{Lemma4}, (and indirectly
illustrate the importance of Corollary~\ref{Cor1}), we compare the time taken
with and without applying the results of Lemma~\ref{Lemma4}. To do so, a
simulated data set consisting of 200 temporal observations from the
hierarchical model in (\ref{eq:meas_DFM}) and (\ref{eq:state_DFM}) is
constructed for specific numbers of time series. In each case, we
consider a two component model, where the common components are
specified with the following dynamics,
\begin{align}
f_{i,t+1} & =  \mu_{i,t+1}+\rho_{i}\left(f_{i,t}-\mu_{i,t}\right)+\sigma_{f,i}\zeta_{i,t},\quad
\mu_{i,t+1}  = \mu_{i,t}+\sigma_{f,i}K_{i,t}^{\mu}\xi_{i,t}.\label{eq:common_fac_i_cc}
\end{align}

\begin{table}
	\centering
{\scriptsize }%
\begin{tabular}{lcccccccccc}
{\scriptsize Number of time series } & {\scriptsize 5 } & {\scriptsize 10 } & {\scriptsize 50 } & {\scriptsize 100 } & {\scriptsize 500 } & {\scriptsize 1000 } & {\scriptsize 5000 } & {\scriptsize 10000 } & {\scriptsize 100000 } & {\scriptsize 500000}\tabularnewline
{\scriptsize Ignores Lemma~\ref{Lemma4} } & {\scriptsize 30 } & {\scriptsize 43 } & {\scriptsize 515 } & {\scriptsize 1407 } & {\scriptsize $27978^{*}$} & {\scriptsize $149831^{*}$} & {\scriptsize $12788430^{*}$} & {\scriptsize $95654412^{*}$} & {\scriptsize NA } & {\scriptsize NA}\tabularnewline
{\scriptsize Uses Lemma 4 } & {\scriptsize 24 } & {\scriptsize 24 } & {\scriptsize 24 } & {\scriptsize 24 } & {\scriptsize 25 } & {\scriptsize 25 } & {\scriptsize 28 } & {\scriptsize 29 } & {\scriptsize 68 } & {\scriptsize 239}\tabularnewline
{\scriptsize Relative speed up } & {\scriptsize 1.25 } & {\scriptsize 1.7 } & {\scriptsize 21 } & {\scriptsize 59 } & {\scriptsize $1119^{*}$} & {\scriptsize $5993^{*}$} & {\scriptsize $456730^{*}$} & {\scriptsize $3298428^{*}$} & {\scriptsize NA } & {\scriptsize NA}\tabularnewline
 &  &  &  &  &  &  &  &  &  & \tabularnewline
\end{tabular}\caption{The table reports the observed or estimated$^{*}$ time (in seconds)
	running 1000 calls of Algorithm~\ref{alg:Sample_K}, both when Lemma~\ref{Lemma4}
	is employed and when it is not, for different sized data sets. We
report the estimated time when it is impractical to run 1000 iterations
for a given data set. In this case fewer iterations are used and the
timings from the reduced run are used to estimate the time taken for
1000 iterations.\label{tab:Lemma_4_timings}}
\end{table}

Table~\ref{tab:Lemma_4_timings} summarizes the computational expense
of using Algorithm~\ref{alg:Sample_K}, both when Lemma~\ref{Lemma4} is used and
when it is not. The first row of the table lists the number of time
series in the analysis. The second and third rows report the time
(note that we report estimated time when it is not practical in all
cases to run 1000 iterations, so fewer iterations are used and the
timings from the reduced run are used to estimate the time taken for
1000 iterations) in seconds for 1000 calls of
Algorithm~\ref{alg:Sample_K}, ignoring Lemma~\ref{Lemma4} and taking advantage of
Lemma~\ref{Lemma4}, respectively. The fourth row reports the relative speed up
that is achieved when taking advantage of Lemma~\ref{Lemma4}. All timings are
done using a Linux operating system, with a 3.2GHz Intel Core i7 processor, with 24
Gigabytes of RAM\@. All computation is done using a combination of the
Python and Fortran programming languages.

The table shows that the savings that arise from Lemma~\ref{Lemma4} are
particularly dramatic as the number of time series grows. In fact, it
is essential to use Lemma~\ref{Lemma4} for any application of large space-time
data sets, as the time difference between using and not using the
lemma can range from a few minutes, to waiting perhaps a few years.

\section{Analysis of Simulated Data\label{sec:Analysis-of-Simulated}}

To critically evaluate the methodology, we first analyze a simulated
data set, consisting of 400 time series of length 300 observations.  A
two factor standard DFM is specified where the factors are essentially
of the same form as (\ref{eq:common_factor_i}). The hyperparameters
are set so that $\rho_{1}=0.8,$ $\rho_{2}=0.9$,
$\lambda_{1}=\lambda_{2}=\frac{2\pi}{23},$$\sigma_{f,1}=\sigma_{f,2}=0.5.$
We set $k^{r}=3$, where the regressors are simulated from a standard
normal distribution. We set the regression coefficients to be the same
for each factor and set $\beta_{11}=\beta_{21}=0.8,$
$\beta_{12}=\beta_{22}=0.9$ and $\beta_{31}=\beta_{32}=0.001.$ The
structural breaks are deterministically rather than stochastically
defined, as this is more sensible for the purpose of validation. The
level for the first common factor, $\mu_{t,1},$ is constant except for
a deterministic break that is defined at the $200^{th}$
observation. The second common factor, $\mu_{t,2},$ is defined to be
constant except for deterministic breaks at the
$50^{th},75^{th},100^{th}$ and $150^{th}$ observations. In addition,
the second common factor has a break in the slope at the $240^{th}$
observation. For identification purposes it is assumed that
$\Theta_{i,i}=1.0$ and for all $j>i$ $\Theta_{j,i}=0.0$; see
\citet{Harvey1989} for further details on identification restrictions
for DFMs. To specify the prior on $\bm{\Theta},$ let
$\bm{\theta}_{i}\in\mathbb{R}^{p-i}$ be a vector formed from the
non-deterministic elements in the $i^{h}$ column of $\bm{\Theta}$. The
prior for $\bm{\Theta}$ is
$p\left(\bm{\Theta}\right)=p\left(\bm{\theta}_{1}\right)p\left(\bm{\theta}_{2}\right)\dots
p\left(\bm{\theta}_{k}\right),$ where
\[
p\left(\bm{\theta}_{i}\right)\sim N\left(\bm{0},\kappa_{i}^{-1}\bm{I}\right),
\]
$\kappa$ is the prior precision, and we assume $\kappa_{i},$
for $i\in\mathbb{N}^{k},$ follows a gamma distribution, such that
$
\kappa_{i}\sim G\left(\frac{\nu_{\kappa}}{2},\frac{S\kappa}{2}\right).
$
Note that, for this prior, conditional on the state, $\bm{x},$ the
$p$ equations that make up the measurement equation are independent
and consequently it is straightforward to sample $\bm{\Theta}$ and
$\bm{\kappa}$ from their respective posterior distributions using
standard Bayesian linear regression theory.

\subsection{Prior Specification}

\begin{table}[h]

{\scriptsize }%
\begin{tabular}{ccccccc}
{\scriptsize Parameter } & {\scriptsize Hyperparameters } & {\scriptsize mean } &  & {\scriptsize Parameter } & {\scriptsize Hyperparameters } & {\scriptsize mean}\tabularnewline
{\scriptsize $\left\{ \rho_{i}\right\} _{i\in\mathbb{N}^{k}}$ } & {\scriptsize $\alpha_{\rho}=15$ and $\beta_{\rho}=1.5$ } & {\scriptsize 0.91 } &  & {\scriptsize $\left\{ \eta_{\delta2,i}\right\} _{i\in\mathbb{N}^{k}}$ } & {\scriptsize $\nu_{\delta2}=3$ and $s_{\delta2}=0.4$ } & {\scriptsize 0.5}\tabularnewline
{\scriptsize $\left\{ \sigma_{f,i}\right\} _{i\in\mathbb{N}^{k}}$ } & {\scriptsize $\nu_{f\sigma}=10$ and $s_{f\sigma}=0.1$ } & {\scriptsize 0.1 } &  & {\scriptsize $\left\{ \lambda_{i}\right\} _{i\in\mathbb{N}^{k}}$ } & {\scriptsize $\alpha_{\lambda}=2,$ $\beta_{\lambda}=2,$ $a=0$ and
$b=\frac{4\pi}{23}$ } & {\scriptsize $\frac{2\pi}{23}$}\tabularnewline
{\scriptsize $\left\{ \eta_{\mu1,i}\right\} _{i\in\mathbb{N}^{k}}$ } & {\scriptsize $\nu_{\mu1}=3$ and $s_{\mu1}=30$ } & {\scriptsize 4.4 } &  & {\scriptsize $\left\{ \sigma_{m,i}\right\} _{i\in\mathbb{N}^{p}}$ } & {\scriptsize $\nu_{m}=10$ and $s_{m}=0.1$ } & {\scriptsize 0.1}\tabularnewline
{\scriptsize $\left\{ \eta_{\mu2,i}\right\} _{i\in\mathbb{N}^{k}}$ } & {\scriptsize $\nu_{\mu2}=3$ and $s_{\mu2}=300$ } & {\scriptsize 13.8 } &  & {\scriptsize $\left\{ \beta_{i,j}\right\} _{i\in\mathbb{N}^{k},j\in\mathbb{N}^{k_{r}}}$ } & {\scriptsize $\sigma_{\beta}=3$ } & {\scriptsize 0.0}\tabularnewline
{\scriptsize $\left\{ \eta_{\delta1,i}\right\} _{i\in\mathbb{N}^{k}}$ } & {\scriptsize $\nu_{\delta1}=3$ and $s_{\delta1}=0.1$ } & {\scriptsize 0.25 } &  & {\scriptsize $\left\{ \varpi_{i,j}\right\} _{i\in\mathbb{N}_{+}^{k},j\in\mathbb{N}^{k_{r}}}$ } & {\scriptsize $p_{\varpi}=0.5$ } & {\scriptsize 0.5}\tabularnewline
\end{tabular}{\scriptsize \par}

\caption{\label{tab:Prior_hyperparameters} The table reports the values of the prior
hyperparameters.}
\end{table}

Table~\ref{tab:Prior_hyperparameters} reports the values of the prior
hyperparameters used in the analysis as well as the corresponding
prior means. The prior mean for $\phi_{i}$ and $\sigma_{i},$ for
$i\in\mathbb{N}^{k},$ imply that \emph{a priori} we assume a fairly
high level of persistence and a small signal for each of the common
factors. The values of the prior hyperparameters for $\eta_{\mu1}$
and $\eta_{\mu2},$ for $i\in\mathbb{N}^{k},$ imply that for each
level we allow for breaks of two different sizes. Likewise, the prior
values for $\eta_{\delta1}$ and $\eta_{\delta2},$ for $i\in\mathbb{N}^{k},$
allows for two different sizes of shifts in the slope. The prior for
$\lambda_{i}$ corresponds to a period of 23 observations. This is
the typical number of observations in one year of MODIS data. In addition
we assume that $\nu_{\kappa}=10$ and $S_{\kappa}=0.01.$ The prior
for $\varpi_{i},$ for $i\in\mathbb{N}^{k},$ $j\in\mathbb{N}^{k^{r}}$
implies that \emph{a priori} that the standard first order autoregressive
process and the stochastic cycle are equally probable.
\begin{figure}
\hspace*{-2mm}\includegraphics[trim=200 0 150 0, clip,width=15cm]{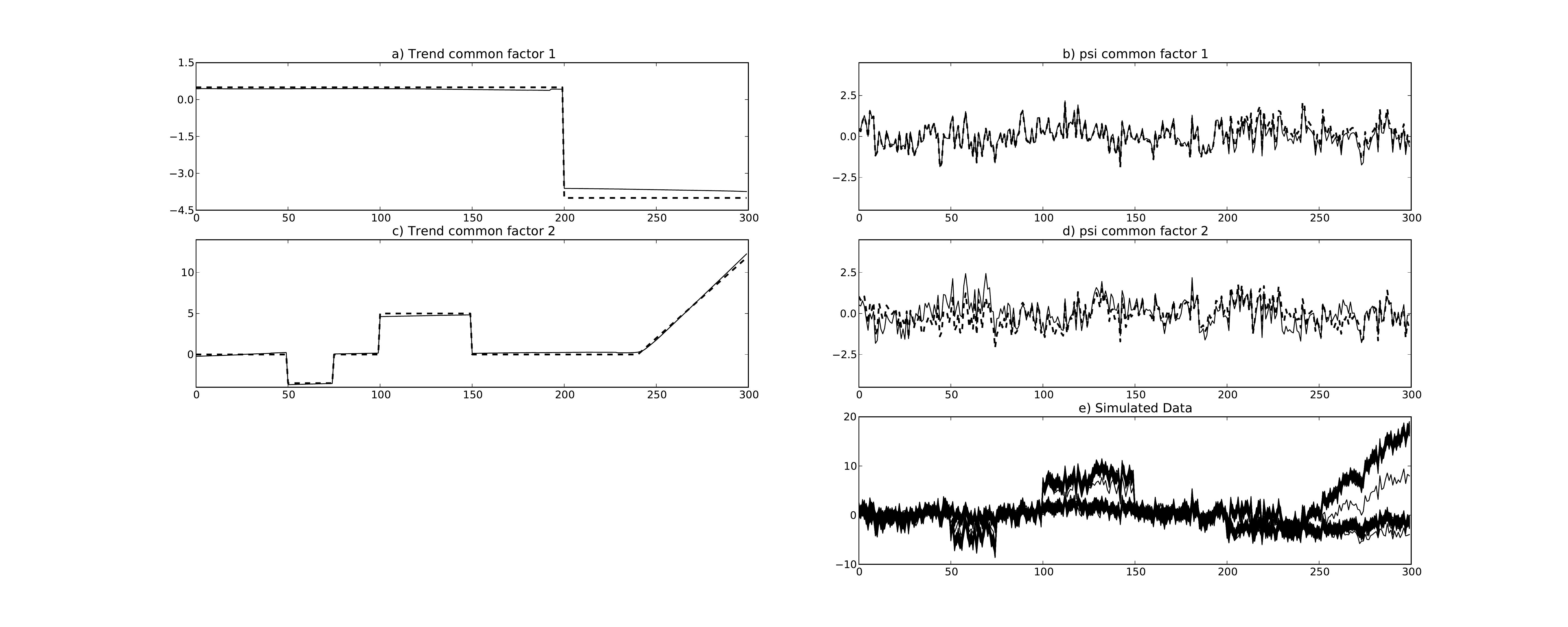}

\caption{\label{fig:Plot_sim_data} a) Plots the estimated (solid line) trend
for factor 1 against the truth (dashed line). b) Plots the estimated
(solid line) for the autoregressive state for factor 1 against the
truth (dashed line). c) Plots the estimated (solid line) trend for
factor 2 against the truth (dashed line). d) Plots the estimated (solid
line) autoregressive state for factor 2, against the truth (dashed
line). e) Plots the simulated data set.}
\end{figure}

Figure~\ref{fig:Plot_sim_data} shows the marginal posterior mean estimates of the trend and autoregressive
component, based on an MCMC analysis, using Algorithm~\ref{alg:MCMC_DFM},
using 5000 iterations, with the
first 1000 discarded.
Panels a) and b) show the marginal posterior mean estimates of the
trend and seasonal component, respectively, for the first common
component, where the seasonal component refers to the cycle
plus regression components. In particular the solid line 
represents the marginal posterior mean estimates and the dashed line is
represents the truth. 
The plots show that the estimates closely follow the truth, and
importantly captures the timing of the changepoint in the level. The
estimates for the second common factor can be seen through Panels c)
and d), and show that the timing of the level shifts have
been accurately captured and further that the shift in slope seems to be
approximately at the correct time. Panel e) plots the simulated data
set.

\section{Modelling change in NDVI from MODIS imagery\label{sec:Modelling-NDVI}}



\begin{figure}
  \centering
  \includegraphics[width=8cm]{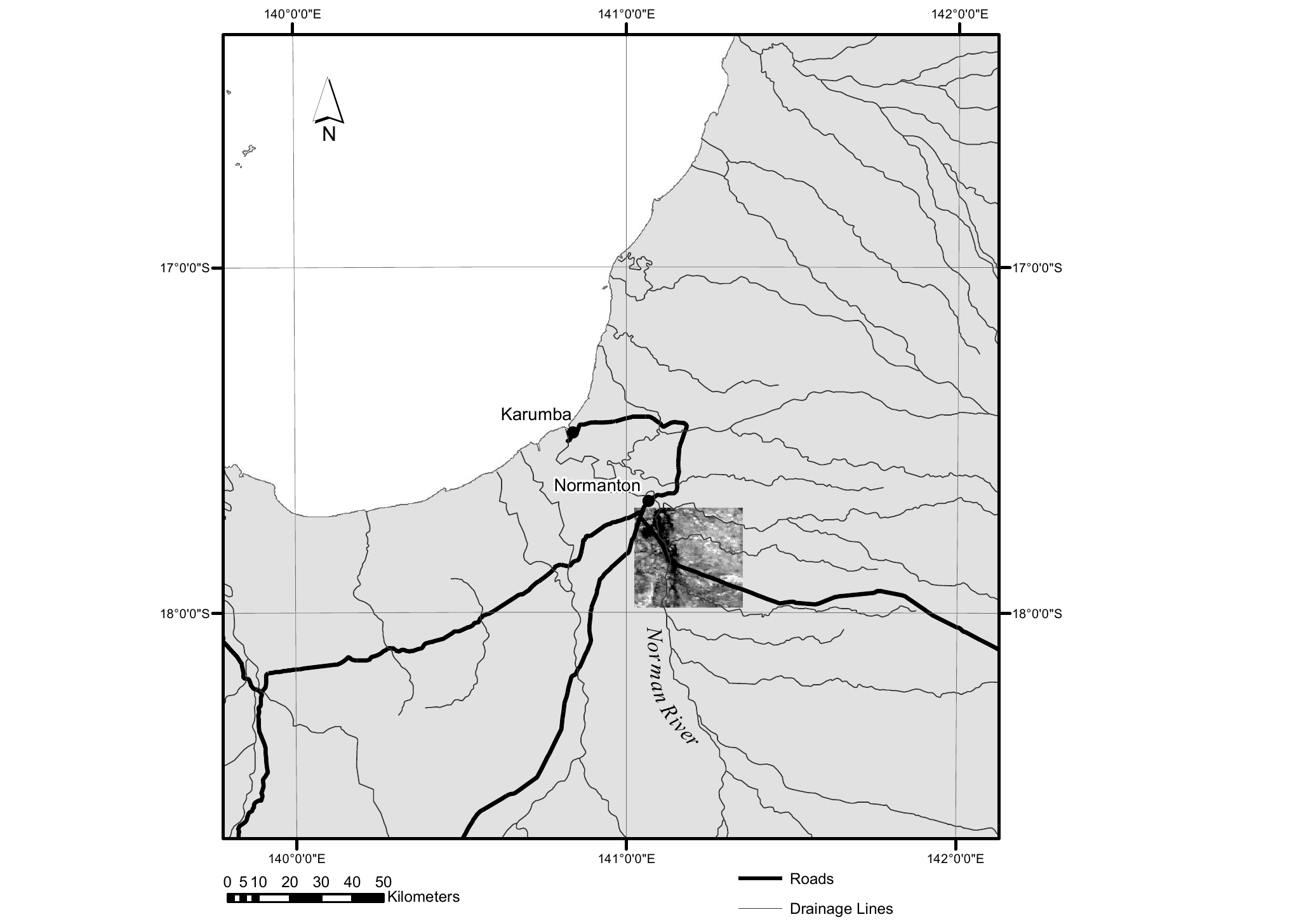}
  \caption{Study site. The subset of MODIS NDVI is an area approximately 35km $\times$32 km south of Normanton, Queensland, Australia.}
  \label{fig:study_site}
\end{figure}

The data set of interest is drawn from an area south of Normanton in
Queensland, Australia (141.187$^{\circ}$ East, 17.843$^{\circ}$
South), and includes part of the Norman River
(Figure~\ref{fig:study_site}).
The data consists of a rectangular array of size $128\times139$ of
NDVI from the MODIS satellite. The pixel size is 250 square meters,
and the area of interest is approximately 35 by 32 kilometers. This product
is available every 16 days, and the data used spans the period from
February 2000 to September 2011. In total, 17653 observations over
space at 268 time points are analysed, which amounts to a total of
4768256 observations.

Plant growth, and hence NDVI, is primarily related to soil moisture
availability, which in turn is related to climatic variables through
precipitation and temperature \citep{Wen2012116,wang2003temporal}. To
model short term variation in NDVI, we used daily climatic data for
the region that are extracted from the SILO climate data bank
\citep{jeffrey01:_using_austr} and averaged over each 16 day interval,
or summed in the case of rainfall. Climatic explanatory variables
considered are maximum temperature, minimum temperature, rainfall,
evaporation, short wave solar radiation for a horizontal surface,
atmospheric water vapour pressure, relative humidity at maximum
temperature, relative humidity at minimum temperature and reference
potential evapotranspiration. Two lags of each regressor are included
as explanatory variables in the model and as a consequence the
regressors can only really be expected to capture relatively short
term seasonal information. We expect any longer term seasonal impact
to feed into the trend.

\begin{figure}[H]
\includegraphics[width=13cm]{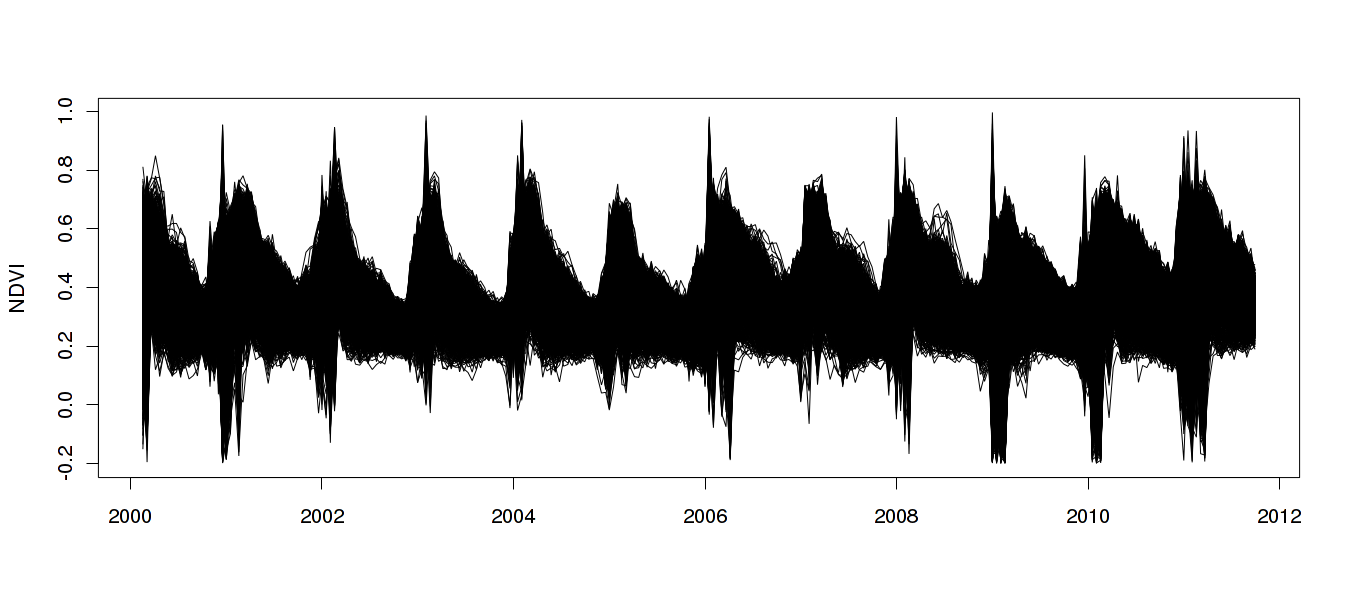}
\caption{Time series plot of the MODIS data set.\label{fig:TS_plot}}
\end{figure}

Figure~\ref{fig:TS_plot} is a time series plot of NDVI for the data set of
interest. The plot shows that the data is seasonal with complex dynamics, but
without any clear evidence of structural breaks.  For the analysis, we use an
EOF basis for $\bm{\Theta},$ where $k$ is set so that the $k^{th}$ component
explains at least one percent of the variation in the data.  The specifications
of the priors remain the same as for the analysis of simulated data, and is thus
described in Table~\ref{tab:Prior_hyperparameters}.

\begin{table}[h]
\begin{tabular}{ccccccccc}
Parameter  & Mean  & Std  & IF  &  & Parameter  & Mean  & Std  & IF\tabularnewline
$\rho_{1}$  & 0.95  & 0.12  & 5.8  &  & $\lambda_{3}$  & 0.25  & 0.03  & 7.81\tabularnewline
$\rho_{2}$  & 0.64  & 0.07  & 7.08  &  & $\lambda_{3}$  & 0.26  & 0.03  & 8.98\tabularnewline
$\rho_{3}$  & 0.68  & 0.06  & 6.31  &  & $\sigma_{f,1}$  & 0.04  & 0.001  & 7.25\tabularnewline
$\rho_{4}$  & 0.58  & 0.08  & 7.31  &  & $\sigma_{f,2}$  & 0.02  & 0.001  & 8.81\tabularnewline
$\lambda_{1}$  & 0.26  & 0.02  & 6.78  &  & $\sigma_{f,3}$  & 0.022  & 0.001  & 6.67\tabularnewline
$\lambda_{2}$  & 0.26  & 0.03  & 8.4  &  & $\sigma_{f,4}$  & 0.03  & 0.001  & 6.7\tabularnewline
\end{tabular}\caption{\label{tab:MCMC_output}The table reports the marginal posterior mean (Mean),
marginal posterior standard deviation (Std) and the Inefficiency factor
(IF), for each of the specified parameters.}
\end{table}

The MCMC analysis using Algorithm~\ref{alg:MCMC_DFM} is run for 5000 iterations, with the first one
thousand iterations are discarded as burnin.
Table~\ref{tab:MCMC_output} reports the estimated output for some of
the parameters from the MCMC analysis of the MODIS data set. It is
clear that for each factor there is a moderate to high level of
persistence in the stochastic cycle and for each case the estimated
period of the cycle is close to a year, which can be expected.
Note that for a period of one year we expect $\lambda=0.26.$
From the inefficiency factors it is evident that the MCMC estimates
are extremely efficient, and in fact for all, of the nearly 18000
parameters, are smaller than nine; see \citet{ChibGreenberg1996} for
further details on inefficiency factors.

\begin{figure}[h]
\hspace*{-2mm}\includegraphics[width=15cm]{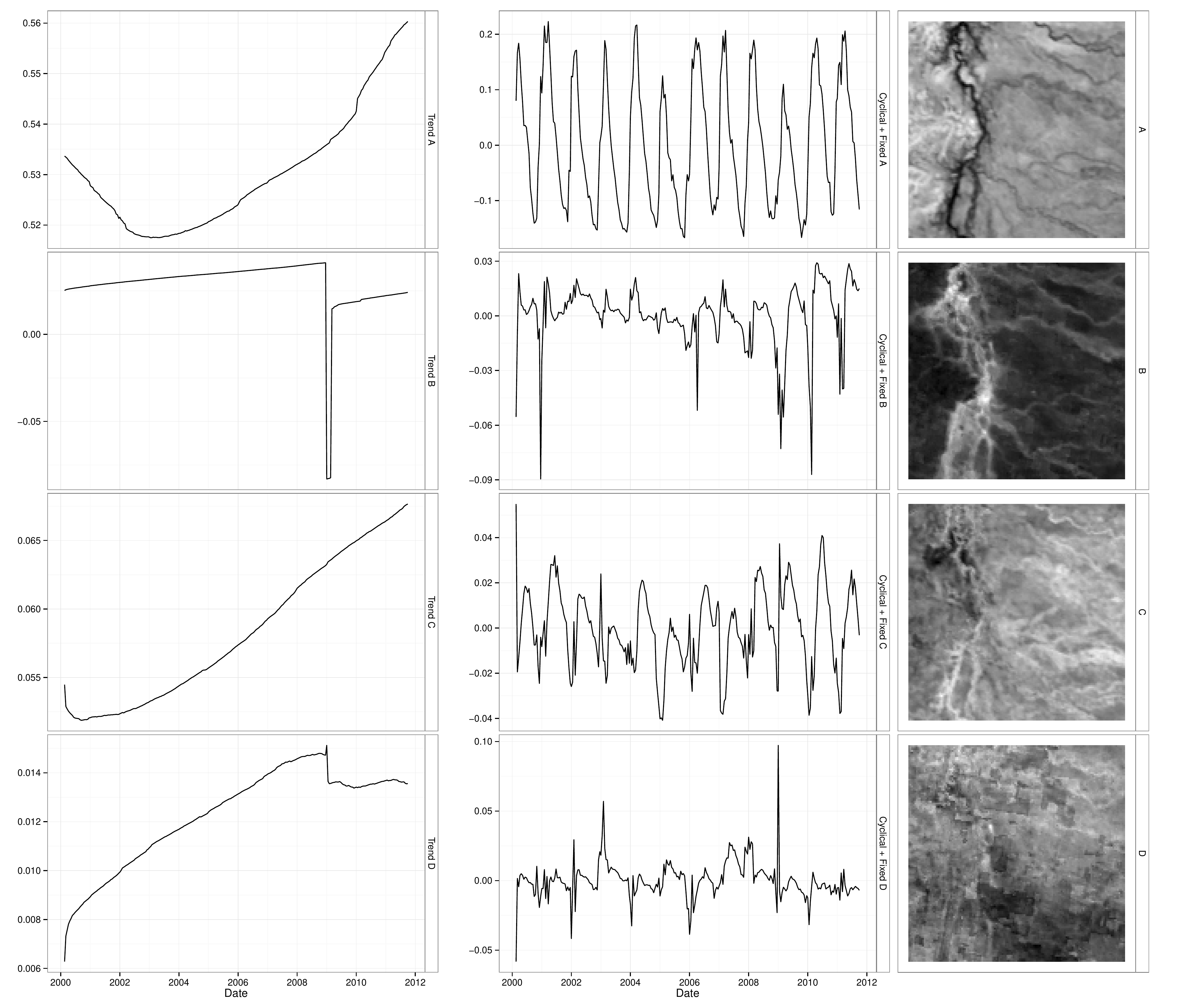}
\caption{\label{fig:plot_of_trend}Plots of the estimated trend, seasonal and
spatial structure for each of the four components.}
\end{figure}

Figure~\ref{fig:plot_of_trend} plots the trend, seasonal component and
the spatial structure in the data, implied by the EOF bases,
respectively, for each of the four components. The trend is 
the marginal posterior mean estimate of $\mu_{i,t},$ for
$t=1,2,\dots,n,$ and can be interpreted as the longer term trend in
the data. The seasonal component is the marginal posterior mean
estimate of $\psi_{i,t}+\bm{w}_{t-1}\bm{\beta}_{i}$ for
$t=2,3,\dots,n$ and for $t=1$ is $\bm{w}_{0}\bm{\beta}_{i}.$
The image plots highlight where each particular component has most
impact. Essentially the lighter the area the more influence the
particular trend has over the data of the corresponding region. Visual
inspection of the image plot for the first component, (A), reveals
that the corresponding common component in the data is least
influential where the water feature is present. For this region,
examination of the trend, suggests that NDVI was initially decreasing, but
has recovered in more recent years. A moderate El Ni\~no in 2002/2003
resulted in well below average rainfall in this area, and the impact
on vegetation is clear in the trend. La Ni\~na events in 2007/2008 and
again in 2008/2009 produced above average rainfall, and, as shown in
the trend, resulted in a general increase in vegetation in the
region. From the image plot we can see that the second component, (B),
is most influential in areas close to the river and its
tributaries. Structural change is clearly present in the trend for this
component. In particular, the trend suggests a substantial drop in NDVI
in 2009, corresponding to a known period of prolonged inundation in
the region. While periodic inundation is common in much of the lower
lying areas in the region, the floods in 2008/2009 were unusual in
that areas were under water for a much longer period. Visual
interpretation of the trend suggests that for this region the level of
NDVI does not immediately recover to previous levels, adding weight to
the theory that the prolonged period of inundation caused long term
damage to the vegetation. The third and fourth components are less
interesting. The calculation of the EOF bases suggest that they
account for far less variation in the data. Arguably, component three,
(C), seems to have most influence in the outer tributaries of the
water system. The trend suggests an overall increase of NDVI over time
for this region, at least above and beyond that of what is explained
by components (A) and (B). The fourth component, (D), arguably shows an
increase in NDVI over time as well, up until the point of the
inundation, where it drops and flattens off. It is also interesting
to note that the seasonal pattern is most regular away from the river
and tributaries. This is not unexpected as this region is
less susceptible to flooding, in which case we can expect a more
uniform response to climatic factors.

\section{Conclusions\label{sec:Conclusions}}

This article introduces a Bayesian methodology for the detection of
structural change in multivariate time series and space-time data.
Remotely sensed data is used in the analysis of the Gulf Plains
bioregion, where, using the proposed methodology, we found evidence of
structural change in a region that had been inundated for an extended
period of time. Areas most affected by the 2009 flood have not
recovered to pre-flood levels in over two and a half years.




This research was partially supported by Australian Research
Council linkage grant, LP100100565. The research of Robert Kohn was
partially supported by Australian Research Council grant
DP066706. All computation was undertaken using the Python
\citet{Python} and Fortran programming languages. We made use of the
libraries, NumPy, SciPy \citet{NumpyScipy}, PyMCMC
\citet{StricklandAlstonDenhamMengersen2012} and PySSM\\
(https://bitbucket.org/christophermarkstrickland/pyssm).  The code
also makes heavy use of BLAS and Lapack though ATLAS\\
(http://math-atlas.sourceforge.net/) and F2PY \citet{Peterson2009}.

\let\normalsize\footnotesize

\appendix

\section{Appendix}

\subsection{Lemmas and Proofs}
\label{Lemma1}
\footnotesize{
\begin{lem}
For $t=1,2,\dots,n,$ the conditional density $\bm{y}^{t+1:n}$ given
$\bm{x}_{t}$ and $\bm{K}$, ignoring terms that are not a function
of $\bm{x}_{t}$ or $\bm{K}$, may be expressed as
\begin{equation}
p\left(\bm{y}^{t+1:n}|\bm{x}_{t},\bm{K},\bm{\omega}\right)\propto\exp\left\{ -\frac{1}{2}\left(\bm{x}_{t}^{T}\bm{\Omega}_{t}\bm{x}_{t}-2\bm{\mu}_{t}^{T}\bm{x}_{t}\right)\right\} ,\label{eq:RC_relation}
\end{equation}
where the terms $\bm{\Omega}_{t}$ and $\bm{\mu}_{t}$ can be computed
through a set of backward recursions. Specifically, the backward recursions
used in sampling $\bm{K}$ for the multivariate conditional state
space model in (\ref{eq:meas}) and (\ref{eq:state}) are computed
by first initializing $\bm{\mu}_{n}=\bm{0}$ and $\bm{\Omega}_{n}=\bm{0},$
where $\bm{\mu}_{t}\in\mathbb{R}^{m}$ and $\bm{\Omega}_{t}\in\mathbb{R}^{m\times m},$
then for $t=n-1,n-2,\dots,1,$ first computing
\allowdisplaybreaks
\begin{align*}
\bm{J}_{t+1} & =  \bm{H}_{t+1}\bm{\Gamma}_{t} &
\bm{R}_{t} & =  \bm{J}_{t+1}\bm{J}_{t+1}^{T}+\bm{G}_{t}\bm{G}_{t}^{T} \\
\bm{L}_{t+1} & =  \bm{\Omega}_{t+1}\bm{C}_{t+1}\bm{D}_{t+1}^{-1}\bm{C}_{t+1}^{T} &
\bm{N}_{t+1} & =  \bm{\Gamma}_{t+1}\left(\bm{I}-\bm{J}_{t+1}^{T}\bm{R}_{t+1}^{-1}\bm{J}_{t+1}\right)\bm{\Gamma}_{t+1}^{T} \\
\bm{C}_{t+1} & =  \bm{N}_{t+1}^{\frac{1}{2}} &
\bm{D}_{t+1} & =  \bm{I}+\bm{C}_{t+1}^{T}\bm{\Omega}_{t+1}\bm{C}_{t+1}\addtocounter{equation}{1}\tag{\theequation} \label{eq:Reduced_Conditionals} \\
\bm{B}_{t+1} & =  \bm{R}_{t+1}^{-1}\bm{J}_{t+1}\bm{\Gamma}_{t+1}^{T}&
\bm{E}_{t+1} & =  \bm{I}-\bm{B}_{t+1}^{T}\bm{H}_{t+1} \\
  \bm{K}_{t} & =  \bm{I}-\bm{L}_{t} &
\bm{M}_{t+1} & =  \bm{R}_{t+1}^{-1}\bm{H}_{t+1}\bm{F}_{t} \\
\bm{S}_{t+1} & =  \bm{K}_{t+1}\bm{\Omega}_{t+1} &
\bm{q}_{t+1} & =  \bm{\Omega}_{t+1}\left(\bm{E}_{t+1}\bm{h}_{t}+\bm{B}_{t+1}^{T}\left(\bm{y}_{t+1}-\bm{g}_{t+1}\right)\right) \\
\bm{A}_{t+1} & =  \bm{E}_{t+1}\bm{F}_{t} \\
\end{align*}
where $\bm{J}_{t},\bm{L}_{t},\bm{R}_{t}\in\mathbb{R}^{p\times p},$
$\bm{B}_{t},\bm{M}_{t}\in\mathbb{R}^{p\times m},$$\bm{N}_{t},\bm{C}_{t},\bm{D}_{t},\bm{E}_{t},\bm{A}_{t},\bm{S}_{t}\in\mathbb{R}^{m\times m}$
and $\bm{q}_{t}\in\mathbb{R}^{m}$ and then computing

\begin{align}
\bm{\Omega}_{t} & =\bm{A}{}_{t+1}^{T}\bm{\Omega}_{t+1}\bm{S}_{t+1}+\bm{M}_{t+1}^{T}\bm{H}_{t+1}\bm{F}_{t}\nonumber \\
\bm{\mu}_{t} & =\bm{A}{}_{t+1}\bm{K}_{t+1}\left(\bm{\mu}_{t+1}-\bm{q}_{t+1}\right)+
 \bm{M}_{t+1}^{T}\left(\bm{y}_{t+1}-\bm{g}_{t+1}-\bm{H}_{t+1}\bm{h}_{t+1}\right).\label{eq:Reduced_Conditionals_2}
\end{align}

\end{lem}
}

\begin{proof}
{\footnotesize To derive the set of equations in (\ref{eq:Reduced_Conditionals})
and (\ref{eq:Reduced_Conditionals_2}), first define $\bm{r}_{t+1},$
such that
\begin{eqnarray*}
\bm{\varepsilon}_{t+1} & = & \bm{y}_{t+1}-\mathbb{E}\left(\bm{y}_{t+1}|\bm{x}_{t},\bm{K}^{1,t+1}\right)
  =  \bm{y}_{t+1}-\bm{g}_{t+1}-\bm{H}_{t+1}\left(\bm{h}_{t}+\bm{F}_{t}\bm{x}_{t}\right)
  =  \bm{J}_{t+1}\bm{u}_{t}+\bm{G}_{t+1}\bm{u}_{t+1},
\end{eqnarray*}
 where $\bm{J}_{t+1}=\bm{H}_{t+1}\bm{\Gamma}_{t},$ and define $\bm{R}_{t+1}$
as
$\bm{R}_{t+1}  =  var\left(\bm{y}_{t+1}|\bm{x}_{t},\bm{K}^{1,t+1}\right)
  =  \bm{J}_{t+1}\bm{J}_{t+1}^{T}+\bm{G}_{t+1}\bm{G}_{t+1}^{T}.$

Noting that $\mathbb{E}\left(\bm{x}_{t+1}|\bm{x}_{t},\bm{y}_{t+1},\bm{K}^{1,t+1}\right)=\mathbb{E}\left(\bm{x}_{t+1}|\bm{x}_{t},\bm{\varepsilon}_{t+1},\bm{K}^{1,t+1}\right),$
$Cov\left(\bm{\varepsilon}_{t},\bm{x}_{t}\right)=\bm{0}$ and $\mathbb{E}\left(\bm{\varepsilon}_{t}\right)=\bm{0},$
it then follows}{\footnotesize \par}
{\footnotesize
\begin{eqnarray*}
\mathbb{E}\left(\bm{x}_{t+1}|\bm{x}_{t},\bm{\varepsilon}_{t+1},\bm{K}\right) & = & \mathbb{E}\left(\bm{x}_{t+1}|\bm{x}_{t},\bm{K}\right)+Cov\left(\bm{x}_{t+1}\bm{\eta}_{t+1}|\bm{K}\right)\bm{R}_{t+1}^{-1}\bm{\eta}_{t+1}\\
 & = & \bm{E}_{t+1}\bm{h}_{t}-\bm{B}_{t+1}^{T}\bm{g}_{t+1}+\bm{B}_{t+1}^{T}\bm{y}_{t+1}+\bm{E}_{t+1}\bm{F}_{t}\bm{x}_{t}\\
 & = & \bm{a}_{t+1}+\bm{A}_{t+1}\bm{x}_{t}+\bm{B}_{t+1}\bm{y}_{t+1},
\end{eqnarray*}
where
\begin{eqnarray*}
	\bm{E}_{t+1} & = & \bm{I}-\bm{B}_{t+1}^{T}\bm{H}_{t+1},\text{     }
\bm{a}_{t+1}  =  \bm{E}_{t+1}\bm{h}_{t}-\bm{B}_{t+1}^{T}\bm{g}_{t+1}\\
\bm{A}_{t+1} & = & \bm{E}_{t+1}\bm{F}_{t}, \text{     }
\bm{B}_{t+1}  =  \bm{R}_{t+1}^{-1}\bm{J}_{t+1}\bm{\Gamma}_{t}^{T}.
\end{eqnarray*}
Let
$\bm{N}_{t+1}=var\left(\bm{x}_{t+1}|\bm{x}_{t},\bm{y}_{t+1},\bm{K}\right).$
Then
$\bm{N}_{t+1}=\bm{\Gamma}_{t}\left(I-\bm{J}_{t+1}^{T}\bm{R}_{t+1}^{-1}\bm{J}_{t+1}\right)\bm{\Gamma}_{t}^{T}.$
Let $\bm{C}_{t+1}=\bm{N}_{t+1}^{1/2},$ where $\bm{N}_{t+1}^{1/2}$
is defined as
$\bm{N}_{t+1}=\left(\bm{N}_{t+1}^{1/2}\right)\left(\bm{N}_{t+1}^{1/2}\right)^{T}.$
Then we can express $\bm{x}_{t+1}$ as}{\footnotesize \par}

{\footnotesize
$\bm{x}_{t+1}=\bm{a}_{t+1}+\bm{A}_{t+1}\bm{x}_{t}+\bm{B}_{t+1}^{T}\bm{y}_{t+1}+\bm{C}_{t+1}\bm{z}_{t+1},$
where $\bm{z}_{t+1}\sim N\left(\bm{0},\bm{I}\right)$ and is independent
of $\bm{x}_{t}$ and $\bm{y}_{t+1}$ (conditional on $\bm{K}).$ We can factor}{\footnotesize \par}

{\footnotesize
$p\left(\bm{y}^{t+1,n}|\bm{x}_{t},\bm{K}\right)=p\left(\bm{y}^{t+2,n}|\bm{y}_{t+1},\bm{x}_{t},\bm{K}\right)\times
p\left(\bm{y}_{t+1}|\bm{x}_{j},\bm{K}\right)$,

where $p\left(\bm{y}^{t+2,n}|\bm{y}_{t+1},\bm{x}_{t},\bm{K}\right)=\int p\left(\bm{y}^{t+2,n}|\bm{x}_{t+1},\bm{K}\right)p\left(\bm{z}_{t+1}|\bm{K}\right)d\bm{z}_{t+1}.$
Using the form of (\ref{eq:RC_relation}) it follows that
\begin{eqnarray*}
p\left(\bm{y}^{t+2,n}|\bm{y}_{t+1},\bm{x}_{t},\bm{K}\right) & = & \int p\left(\bm{y}^{t+2,n}|\bm{x}_{t+1},\bm{K}\right)p\left(\bm{z}_{t+1}|\bm{K}\right)d\bm{z}_{t+1}\\
 & \propto & \exp\left\{ -\frac{1}{2}\left[\bm{x}_{t}^{T}\left(\bm{A}_{t+1}^{T}\bm{\Omega}_{t+1}\bm{A}_{t+1}-\bm{A}_{t+1}^{T}\bm{\Omega}_{t+1}\bm{C}_{t+1}\bm{D}_{t+1}^{-1}\bm{C}_{t+1}^{T}\bm{\Omega}_{t+1}\bm{A}_{t+1}\right)\bm{x}_{j}\right.\right.\\
 & - & \left.\left.2\bm{x}_{j}^{T}\bm{A}_{t+1}^{T}\left(\left(\bm{I}-\bm{\Omega}_{t+1}\bm{C}_{t+1}\bm{D}_{t+1}^{-1}\bm{C}_{t+1}^{T}\right)\mbox{\ensuremath{\left(\bm{\mu}_{t+1}-\bm{\Omega}_{t+1}\left(\bm{a}_{t+1}+\bm{B}_{t+1}^{T}\bm{y}_{t+1}\right)\right)}}\right)\right]\right\} ,
\end{eqnarray*}
}{\footnotesize \par}

{\footnotesize where
$\bm{D}_{t+1}=\bm{I}+\bm{C}_{t+1}^{T}\bm{\Omega}_{t+1}\bm{C}_{t+1}.$
 Combining $p\left(\bm{y}^{t+2,n}|\bm{y}_{t+1},\bm{x}_{t},\bm{K}\right)$
with $p\left(\bm{y}^{t+2,n}|\bm{y}_{t+1},\bm{x}_{t},\bm{K}\right)$,
where}

{\footnotesize
\[
p\left(\bm{y}_{t+1}|\bm{x}_{t},\bm{K}\right)\propto\exp\left\{ -\frac{1}{2}\left(\bm{y}_{t}-\bm{g}_{t+1}-\bm{H}_{t+1}\left(\bm{h}_{t}+\bm{F}_{t}\bm{x}_{t}\right)\right)\bm{R}^{-1}\left(\bm{y}_{t}-\bm{g}_{t+1}-\bm{H}_{t+1}\left(\bm{h}_{t}+\bm{F}_{t}\bm{x}_{t}\right)\right)^{T}\right\} ,
\]
 and completing the square then it follows that}{\footnotesize \par}

{\footnotesize
$\bm{\Omega}_{t}=\bm{A}{}_{t+1}^{T}\bm{\Omega}_{t+1}\bm{S}_{t+1}+\bm{M}_{t+1}^{T}\bm{H}_{t+1}\bm{F}_{t},$
 with $\bm{S}_{t+1}=\bm{K}_{t+1}\bm{\Omega}_{t+1},$ where $\bm{K}_{t+1}=\bm{I}-\bm{L}_{t},$
$\bm{L}_{t}=\bm{\Omega}_{t+1}\bm{C}_{t+1}\bm{D}_{t+1}^{-1}\bm{C}_{t+1}^{T}$
and $\bm{M}_{t+1}=\bm{R}_{t+1}^{-1}\bm{H}_{t+1}\bm{F}_{t+1}.$ Further,
\[
\bm{\mu}_{t}=\bm{A}{}_{t+1}\bm{K}_{t+1}\left(\bm{\mu}_{t+1}-\bm{q}_{t+1}\right)+\bm{M}_{t+1}^{T}\left(\bm{y}_{t+1}-\bm{g}_{t+1}-\bm{H}_{t+1}\bm{h}_{t+1}\right).
\]
 }{\footnotesize \par}
 \end{proof}

\footnotesize{
\begin{lem}
\label{Lemma2}
The conditional density of $\bm{y}_{t}$ given $\bm{y}^{1:t-1}$ and
$\bm{K}^{1:t}$ may be expressed as
\[
p\left(\bm{y}_{t}|\bm{y}^{1:t-1},\bm{K}^{1:t},\bm{\theta}\right)\propto\left|\bm{R}_{t}\right|^{-\frac{1}{2}}\exp\left\{ -\frac{1}{2}\left(\bm{\eta}_{t}^{T}\bm{R}_{t}^{-1}\bm{\eta}_{t}\right)\right\} ,
\]
where the quantities $\bm{\eta}_{t}\in\mathbb{R}^{p}$ and $\bm{R}_{t}\in\mathbb{R}^{p\times p}$
are calculated using the following recursion, with $\bm{m}_{1}$ and
$\bm{V}_{1}$ obtained from the prior in (\ref{eq:init_state}),
\begin{align*}
	\bm{r}_{t} & =  \bm{y}_{t}-\bm{g}_{t}-\bm{H}_{t}\bm{m}_{t} &
\bm{m}_{t} & =  \bm{h}_{t-1}+\bm{F}_{t-1}\bm{m}_{t-1|t-1}\\
\bm{R}_{t} & =  \bm{H}_{t}\bm{V}_{t}\bm{H}_{t}^{T}+\bm{G}_{t}\bm{G}_{t}^{T} &
\bm{J}_{t} & =  \bm{R}_{t}^{-1}\bm{H}_{t}\bm{V}_{t}\\
\bm{V}_{t} & =  \bm{F}_{t-1}\bm{V}_{t-1|t-1}\bm{F}_{t-1}^{T}+\bm{\Gamma}_{t-1}\bm{\Gamma}_{t-1}^{T} &
\bm{V}_{t|t} & =  \bm{V}_{t}-\bm{J}_{t}^{T}\bm{R}_{t}\bm{J}_{t}\\
\bm{m}_{t|t} & =  \bm{m}_{t}+\bm{J}_{t}^{T}\bm{\eta}_{t}.\\
\end{align*}

\end{lem}
}
\begin{proof}
{\footnotesize It is straightforward to verify that
$\bm{r}_{t} = \bm{y}_{t}-\mathbb{E}\left(\bm{y}_{t}|\bm{y}^{1,t-1},\bm{K}^{1,t}\right) = \bm{g}_{t}+\bm{H}_{t}\bm{m}_{t},$
where
$\bm{m}_{t} = \mathbb{E}\left(\bm{x}_{t}|\bm{y}^{1,t-1},\bm{K}^{1,t}\right) =  \bm{f}_{t-1}+\bm{F}_{t-1}\bm{m}_{t-1|t-1}.$
}{\footnotesize}
Furthermore,
{\footnotesize
$\bm{R}_{t}  = Cov\left(\bm{\eta}_{t}|\bm{y}^{1,t-1},\bm{K}\right) =  \mathbb{E}\left(\bm{\eta}_{t}\bm{\eta}_{t}^{T}|\bm{y}^{1,t-1},\bm{K}\right)
  = \bm{H}_{t}\bm{V}_{t}\bm{H}_{t}^{T}+\bm{G}_{t}\bm{G}_{t}^{T},$
where
$\bm{V}_{t} = Cov\left(\bm{x}_{t}|\bm{y}^{1,t-1},\bm{K}^{1,t}\right)
  =  \bm{F}_{t-1}\bm{V}_{t-1|t-1}\bm{F}_{t-1}^{T}+\bm{\Gamma}_{t-1}\bm{\Gamma}_{t-1}^{T}.$
}

{\footnotesize
\begin{eqnarray*}
\bm{m}_{t|t} & = & \mathbb{E}\left(\bm{x}_{t}|\bm{y}^{1,t},\bm{K}\right) = \mathbb{E}\left(\bm{x}_{t}|\bm{y}^{1,t-1},\bm{K}\right)+Cov\left(\bm{x}_{t},\bm{\eta}_{t}|\bm{y}^{1,t-1},\bm{K}\right)Cov\left(\bm{\eta}_{t}|\bm{y}^{1,t-1},\bm{K}\right)^{-1}\bm{\eta}_{t}\\
 & = & \bm{m}_{t-1}+\bm{J}_{t}^{T}\bm{\eta}_{t},
\end{eqnarray*}
where
$\bm{J}_{t}=\bm{R}_{t}^{-1}\bm{H}_{t}\bm{V}_{t|t},$
}
{\footnotesize with
$\bm{V}_{t|t}=\bm{V}_{t}-\bm{J}_{t}^{T}\bm{R}_{t}\bm{J}_{t}$
 and}{\footnotesize \par}

{\footnotesize
\begin{eqnarray*}
\bm{V}_{t|t} & = & Cov\left(\bm{x}_{t}|\bm{y}^{1,t},\bm{K}\right)\\
 & = & Cov\left(\bm{x}_{t}|\bm{y}^{1,t-1},\bm{K}\right)-Cov\left(\bm{x}_{t},\bm{\eta}_{t}|\bm{y}^{1,t-1},\bm{K}\right)Cov\left(\bm{\eta}_{t}|\bm{y}^{1,t-1},\bm{K}\right)^{-1}Cov\left(\bm{x}_{t},\bm{\eta}_{t}|\bm{y}^{1,t-1},\bm{K}\right)^{T}\\
 & = & \bm{V}_{t}-\bm{J}_{t}^{T}\bm{R}_{t}\bm{J}_{t}.
\end{eqnarray*}
}{\footnotesize \par}
\end{proof}

\footnotesize{
\begin{lem}
\label{Lemma3}
Factorize $\bm{V}_{t|t}$ as $\bm{T}\bm{T}^{T}$and write $\bm{x}_{t}=\bm{m}_{t}+\bm{T}_{t}\bm{\xi}_{t},$
where $\bm{\xi}_{t}\sim\mathcal{N}\left(\bm{0},\bm{I}\right)$ and
is independent of $\bm{y}^{1:t}.$ It then follows that the conditional
density for $\bm{y}_{t}$ given $\bm{y}^{1:t-1}$ and $\bm{K}^{1:t}$
is
\begin{eqnarray}
p\left(\bm{y}^{t+1:n}|\bm{y}^{1:t},\bm{K}\right) & = & \int p\left(\bm{y}^{t+1:n}|\bm{x}_{t}\right)p\left(\bm{\xi}_{t}|\bm{K}^{1:t}\right)\nonumber \\
 & \propto & \left|\bm{Z}_{t}\right|^{-1/2}\exp\left\{ -\frac{1}{2}\left[\bm{m}_{t}^{T}\left(\bm{\Omega}_{t}\bm{m}_{t}-2\bm{\mu}_{t}\right)-\bm{o}_{t}^{T}\bm{T}_{t}\bm{Z}_{t}^{-1}\bm{T}_{t}^{T}\bm{o}_{t}\right]\right\} ,\label{eq:cond_d_data}
\end{eqnarray}
where $\bm{o}_{t}=\bm{\mu}_{t}-\bm{\Omega}_{t}\bm{m}_{t}$ and
$\bm{Z}_{t}=\bm{T}_{t}^{T}\bm{\Omega}_{t}\bm{T}+\bm{I}$. The proof of this
Lemma follows directly from \citet{GerlachCarterKohn2000}.
\end{lem}
}

\subsubsection*{Proof of Lemma~\ref{Lemma4}}
\begin{proof}
	\label{ProofLemma4}
{\footnotesize To prove Lemma~\ref{Lemma4} holds we need to show that $p\left(\bm{K}_{t}|\bm{y}^{L},\bm{K}_{t\ne s},\bm{\Sigma}^{L},\bm{\omega}\right)\propto p\left(\bm{K}_{t}|\bm{y},\bm{K}_{t\ne s},\bm{\Theta},\bm{\Sigma},\bm{\omega}\right),$
when $\bm{K}_{t}$ only enters through the state equation. We begin
by expressing (\ref{eq:DFM}) as
\begin{equation}
\tilde{\bm{y}}_{t}=\tilde{\bm{\Theta}}\bm{f}_{t}+\tilde{\bm{e}}_{t};\,\,\,\tilde{\bm{e}}_{t}\sim N\left(\bm{0},\bm{I}_{p}\right),\label{eq:Transformed_DFM}
\end{equation}
where $\bm{\tilde{y}}_{t}=\bm{\Sigma}^{1/2}\bm{y}_{t}$ and $\tilde{\bm{\Theta}}=\bm{\Sigma}^{1/2}\bm{\Theta}.$
Next, we decompose $\tilde{\bm{\Theta},}$ using the QR decomposition,
such that
$\bm{\tilde{y}}_{t}=\left[\bm{Q}_{1},\bm{Q}_{2}\right]\left[\begin{array}{c}
\bm{R}\\
\bm{0}
\end{array}\right]\bm{f}_{t}+\tilde{\bm{e}}_{t},$
where $\bm{Q}_{1}\in\mathbb{R}^{p\times k}$ and $\bm{Q}_{2}\in\mathbb{R}^{k\times\left(p-k\right)}$
have orthogonal columns. It follows that
$\left[\begin{array}{c}
\bm{Q}_{1}^{T}\\
\bm{Q}_{2}^{T}
\end{array}\right]\tilde{y}_{t}=\left[\begin{array}{c}
\bm{R}\\
\bm{0}
\end{array}\right]\bm{f}_{t}+\tilde{\bm{e}}_{t}.$
If we define $\left[\begin{array}{c}
\bm{z}_{1}\\
\bm{z}_{2}
\end{array}\right]=\left[\begin{array}{c}
\bm{Q}_{1}^{T}\\
\bm{Q}_{2}^{T}
\end{array}\right]\tilde{\bm{y}}_{t}$ then
$\left[\begin{array}{c}
\bm{z}_{1}\\
\bm{z}_{2}
\end{array}\right]=\left[\begin{array}{c}
\bm{R}f_{t}+\bm{e}_{1,t}\\
\bm{e}_{2,t}
\end{array}\right].$
Clearly,
\begin{equation}
p\left(\bm{K}_{t}|\bm{K}_{s\ne t}|\bm{z}_{1},\bm{\omega},\bm{\Sigma}\right)\propto p\left(\bm{K}_{t}|\bm{K}_{s\ne t}|\bm{z}_{1},\bm{z}_{2},\bm{\omega},\bm{\Sigma}\right).\label{eq:posterior_K_given_z}
\end{equation}
because $\bm{K}_{t}$ only enters through the state
transition equation.
Note that for the transformed measurement equation in (\ref{eq:Transformed_DFM})
\begin{eqnarray}
\bm{y}_{t}^{L} & = & \left(\tilde{\bm{\Theta}}^{T}\tilde{\bm{\Theta}}\right)^{-1}\tilde{\bm{\Theta}}^{T}\tilde{\bm{y}}_{t}\nonumber
  = \left(\bm{R}^{T}\bm{R}\right)^{-1}\left[\begin{array}{cc}
\bm{R}^{T} & \bm{0}\end{array}\right]\left[\begin{array}{c}
\bm{z}_{1,t}\\
\bm{z}_{2,t}
\end{array}\right]\nonumber \\
 & = & \left(\bm{R}^{T}\bm{R}\right)^{-1}\bm{R}^{T}\bm{z}_{1,t}.\label{eq:ytL_re_expressed}
\end{eqnarray}
Lemma~\ref{Lemma4} follows from (\ref{eq:posterior_K_given_z}) and (\ref{eq:ytL_re_expressed}). }{\footnotesize \par}
\end{proof}

\subsection{MCMC Sampling Scheme}
\footnotesize{
\begin{algorithm}[h]
\begin{enumerate}
\item Sample $\bm{K}^{\left(j\right)}$ from $p\left(\bm{K},|\bm{y},\bm{\beta}^{\left(j-1\right)}\mbox{,}\bm{\Theta}^{\left(j-1\right)},\bm{\phi}^{(j-1)},\bm{\sigma}_{f}^{(j-1)},\bm{\lambda}^{\left(j-1\right)},\bm{\Sigma}^{\left(j-1\right)},\bm{\eta}^{\left(j-1\right)},\bm{\varpi}^{\left(j-1\right)}\right),$
where $\bm{\phi}=\left(\phi_{1},\phi_{2},\dots,\phi_{k}\right)$,
$\bm{\sigma}_{f}=\left(\sigma_{f,1},\sigma_{f,2},\dots,\sigma_{f,k}\right),$
$\bm{\lambda}=\left(\lambda_{1},\lambda_{2},\dots,\lambda_{k}\right)$
and $\bm{\varpi}=\left(\varpi_{1},\varpi_{2},\dots,\varpi_{k}\right).$
\item Sample $\bm{x}^{\left(j\right)}$and $\bm{\beta}^{\left(j\right)},$
jointly from $p\left(\bm{x},\bm{\beta}|\bm{y},\bm{K}^{\left(j\right)}\mbox{,}\bm{\Theta}^{\left(j-1\right)},\bm{\phi}^{(j-1)},\bm{\sigma}_{f}^{(j-1)},\bm{\lambda}^{\left(j-1\right)},\bm{\Sigma}^{\left(j-1\right)},\bm{\eta}^{\left(j-1\right)},\bm{\varpi}^{\left(j-1\right)}\right)$
\item Sample $\bm{\eta}^{\left(j\right)}$ from $p\left(\bm{\eta}|\bm{x}^{\left(j\right)},\bm{K}^{\left(j\right)},\bm{\beta}^{\left(j\right)},\bm{\phi}^{(j-1)},\bm{\sigma}_{f}^{(j-1)},\bm{\lambda}^{\left(j-1\right)},\bm{\varpi}^{\left(j-1\right)}\right)$.
\item Sample $\bm{\phi}^{\left(j\right)}$ from $p\left(\bm{\phi}|\bm{y},\bm{K}^{\left(j\right)},\bm{\Theta}^{\left(j-1\right)},\bm{\beta}^{\left(j\right)},\bm{\sigma}_{f}^{(j-1)},\bm{\Sigma}^{\left(j-1\right)},\bm{\eta}^{\left(j\right)}\right)$.
\item Sample $\bm{\sigma}_{f}^{\left(j\right)}$ from $p\left(\bm{\phi}|\bm{y},\bm{K}^{\left(j\right)},\bm{\Theta}^{\left(j-1\right)},\bm{\beta}^{\left(j\right)},\bm{\phi}^{(j)},\bm{\lambda}^{\left(j-1\right)},\Sigma^{\left(j-1\right)},\bm{\eta}^{\left(j\right)}\right)$.
\item Sample $\bm{\lambda}^{\left(j\right)}$ from $p\left(\bm{\lambda}|\bm{y},\bm{K}^{\left(j\right)},\bm{\beta}^{\left(j\right)},\bm{\phi}^{\left(j\right)},\bm{\eta}^{\left(j\right)},\bm{\sigma}_{f}^{\left(j\right)}\right).$
\item Sample $\bm{\varpi}^{\left(j\right)}$ from $p\left(\bm{\varpi}|x^{\left(j\right)},\bm{K}^{\left(j\right)},\bm{\lambda}^{\left(j\right)},\bm{\phi}^{\left(j\right)},\bm{\eta}^{\left(j\right)},\bm{\sigma}_{f}^{\left(j\right)}\right).$
\item Sample $\bm{\Theta}^{\left(j\right)},\bm{\kappa}$ from $p\left(\bm{\Theta}|\bm{y},\bm{x}^{\left(j\right)},\bm{\Sigma}^{\left(j\right)},\bm{\kappa}^{\left(j-1\right)}\right)$.
\item Sample $\bm{\Sigma}^{\left(j\right)}$ from $p\left(\bm{\Sigma}|\bm{y},\bm{x}^{\left(j\right)},\bm{\Theta}^{\left(j\right)}\right).$
\end{enumerate}
\caption{\label{alg:MCMC_DFM}}
\end{algorithm}

Algorithm~\ref{alg:MCMC_DFM} defines the MCMC algorithm for the
hierarchical multivariate time series and space-time model that is
being considered.  Step 1 is unchanged from
Algorithm~\ref{alg:MCMC}. Step 2 only requires a small modification
from Step 2 of Algorithm~\ref{alg:MCMC}. In particular, the algorithm
to sample $\bm{x}$ is augmented to now sample $\bm{x},$ jointly with
$\bm{\beta}$; see \citet{deJongShepard1995} for details on the
modifications required to efficiently jointly sample $\bm{x}$ and
$\bm{\beta}.$ The remaining steps are specific to the hierarchical
model of interest. Step 3 is carried out by sampling each element of
$\bm{\eta}$ from its posterior distribution, which are all inverted
gamma distributions. In Steps 4, 5 and 6 each element of $\bm{\phi},$
$\bm{\sigma}_{f}$ and $\bm{\lambda}$ is sampled individually using
adaptive random walk Metropolis Hastings algorithms; see
\citet{GarthwaiteFanScisson2010} for further details. The sampling is
done marginally of the state, $\bm{x},$ by taking advantage of Corollary~\ref{Cor1}.
In Step 7, we can take advantage of the standard form of the
posterior, and sample each element of $\bm{\varpi}$ which has a closed
form solution. In particular, for each element of $\bm{\varpi}$, the
posterior distribution is a Bernoulli distribution.  In
Step 8, sampling $\bm{\Theta}$ and $\bm{\kappa},$ in the case that
$\bm{\Theta}$ is unknown, depends on its specification so details are
given in the relevant application sections. Step 9 is straightforward
as the diagonal elements in $\bm{\Sigma}$ are conditionally
independent with inverted gamma posterior distributions.\emph{ }
}
{\footnotesize{} \bibliographystyle{apalike}
\bibliography{reduced_cond}
}
\end{document}